\documentclass[%
 reprint,
 amsmath,amssymb,
 aps,
]{revtex4-2}

\usepackage{graphicx}
\usepackage{dcolumn}
\usepackage{bm}
\usepackage{hyperref}
\usepackage[mathlines]{lineno}
\usepackage{color}
\usepackage{amsmath}
\usepackage[normalem]{ulem}

\begin{document}

\preprint{APS/123-QED}

\title{Dual phase transitions in a 1D lattice with PT-symmetric Floquet defect}

\author{Zhenzhi Liu, Ke Li, Yanpeng Zhang, Fu Liu}
 \email{fu.liu@xjtu.edu.cn}
\affiliation{Key Laboratory for Physical Electronics and Devices of the Ministry of Education \& Shaanxi Key Lab of Information Photonic Technique, School of Electronic Science and Engineering, Faculty of Electronic and Information Engineering, Xi’an Jiaotong University, Xi’an, 710049, China}


\date{\today}

\begin{abstract}
Systems with non-Hermitian potential or Floquet modulation often result in phase transition related phenomena. In this paper, we study the dual phase transitions in a one-dimensional lattice by introducing a defect containing both Floquet modulation and PT-symmetric potential. In such a configuration, we demonstrate how the gain-loss from PT-symmetry and the control parameters in Floquet modulation adjust the wave dynamic behaviors. When these parameters change, the system will undergo dual phase transitions from an energy-delocalized phase to a localized phase where energy oscillates with time, and then to a PT-symmetry broken phase with energy boost. In particular, we find that the energy oscillations in the second phase is resulted from the beating of two energy oscillations: one is introduced by the PT-symmetric potential and the other is introduced by the Floquet modulation, rather than the field interference of the defect modes. Furthermore, we find that the first phase transition can be non-exist and the second phase transition is affected by the Floquet parameters. Our results reveal the underlying physics of dual phase transitions that occur in simple lattice systems with PT-symmetric Floquet defect, which extends the study of non-Hermitian Floquet systems.

\end{abstract}

\maketitle


\section{\label{sec:level1}Introduction}

Phase transitions, as one of the core research topics in physics, are extensively studied in both classical and quantum systems~\cite{RevModPhys.69.315, PhysRevLett.95.105701, PhysRevLett.124.037801}. In addition to the commonly-known ones between solid, liquid, and gas, typical phase transitions also include metal-insulator transitions~\cite{PhysRevX.12.021067, PhysRevResearch.2.032071}, topological phase transitions~\cite{roushan2014observation, PhysRevLett.117.013902}, Parity-time (PT) symmetric phase transitions~\cite{ruter2010observation, feng2017non, PhysRevA.99.043412}, and so on, where symmetry plays a crucial role. In the study of phase transitions, a critically important concept, i.e., spontaneous symmetry breaking, is introduced. For example, in typical PT-symmetric systems, spontaneous symmetry breaking exists, and the corresponding exceptional point divides the system into PT-symmetric and PT-broken phases~\cite{PhysRevLett.80.5243}. The research on PT-symmetry has expanded the understanding of quantum mechanics, enabling one to predict unique phenomena that do not exist in Hermitian systems~\cite{liu2024floquet, science.abf6873}. The phase transition processes of PT-symmetric systems have been demonstrated in optics and other classical wave systems~\cite{ruter2010observation, feng2017non, shi2016accessing, cao2022fully, PhysRevLett.132.113802}. In addition, PT-symmetry has also been extended to other fields, such as laser devices~\cite{cseker2023single}, non-Hermitian topological insulators~\cite{PhysRevResearch.3.043006}, observation of optical solitons~\cite{wimmer2015observation} and Bloch oscillations~\cite{xu2016experimental}, etc.

\begin{figure*}
\includegraphics[width=0.7\textwidth]{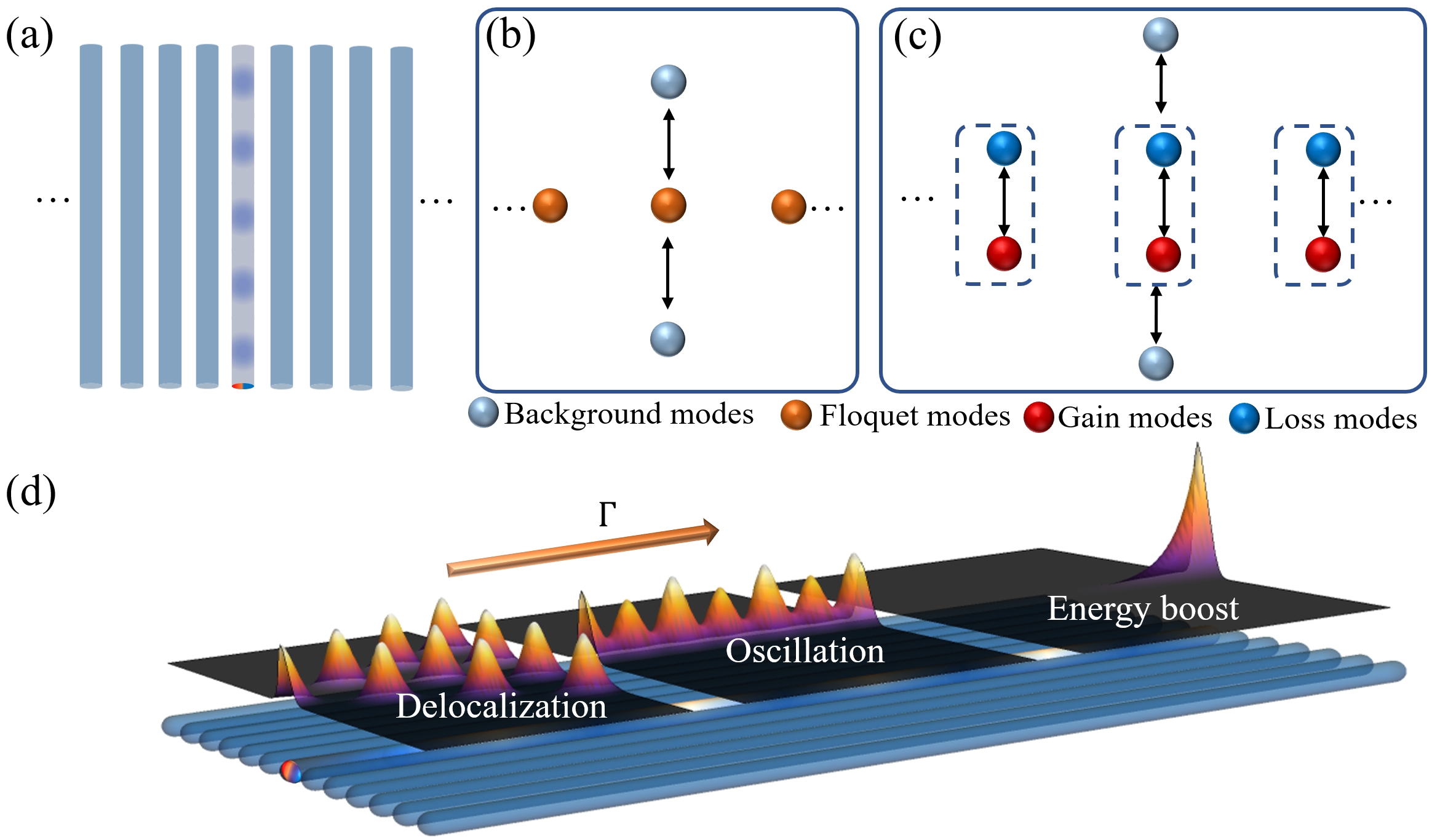}
	\caption{Schematic diagrams of (a) the system, (b, c) modes, and (d) wave dynamic behaviors. (a) The system consists of a PT-symmetric Floquet defect (middle) in a periodic lattice array. The light blue-gray variation in the defect means the Floquet modulation of the potential while the red (blue) button at the bottom indicates the gain (loss). (b, c) The different modes and their couplings in (b) traditional Floquet system without PT-symmetric potential and (c) our proposed system with PT-symmetric Floquet defect. (d) Three different wave dynamic behaviors in the three phases, i.e., delocalization, localized oscillation, and exponential energy boost, of the system under different gain-loss parameter $\Gamma$.}\label{fig:schmatic}
\end{figure*}

Typically, PT-symmetric systems are open systems and do not obey energy conservation. Similarly, time-varying systems, due to the existence of external drivers for time modulation, break the continuous time translation symmetry and therefore they also belong to the category of open systems~\cite{yin2022floquet}. Time-varying or time-modulated systems are widely studied due to their peculiar characteristics~\cite{lin2016photonic, li2023direct, cheng2023multi, PhysRevLett.120.043902, PhysRevLett.128.186802, dong2024quantum}. Among these, a special class of Floquet systems, in which potential changes periodically in time, gained extensive attention from researchers recently~\cite{PhysRevLett.117.090402, RevModPhys.89.011004, geier2021floquet, li2023fractal}. Floquet modulation is also used to control the phase transitions in optics~\cite{yin2022floquet} and condensed matter systems~\cite{PhysRevLett.120.096401, PhysRevLett.130.233801, zhou2023pseudospin,lindner2011floquet, lustig2018topological}. For example, based on the periodic modulation of refractive index or surface impedance, the system can transit from a trivial phase to a photonic time crystal phase~\cite{wang2023metasurface,lyubarov2022amplified, liu2023photonic}. Unlike the photonic crystals with spatially periodic modulation of refractive index, time modulation adds new degrees of freedom for light field manipulation, and various applications including PT-symmetry~\cite{li2020parity}, maser~\cite{jiang2021floquet}, wireless transfer of power~\cite{PhysRevApplied.16.014017}, superradiance lattices~\cite{PhysRevLett.129.273603}, to name a few, are achieved. 

Periodically modulated lattice systems provide an ideal platform for studying the Floquet modulation. Under the paraxial approximation, the wave dynamics in these lattices can be described using the Schrödinger-like equation, where the direction of light propagation is analogous to the time freedom of a quantum system, thus enabling the mimic of quantum systems~\cite{block2014bloch,rechtsman2013photonic, PhysRevResearch.2.043239, PhysRevApplied.20.054005}. There are typically two ways to realize the periodic modulation: one is the helical lattices~\cite{pyrialakos2022bimorphic, rechtsman2013photonic}, and another is the straight lattices with periodically varying refractive index in the propagation direction~\cite{pyrialakos2022bimorphic, PhysRevLett.133.073803, fedorova2019limits}. Moreover, besides the introduction of Floquet modulation in the entire lattice array, Floquet modulation can also be introduced as a defect, as shown in Fig.~\ref{fig:schmatic}(a), where the middle one is under Floquet modulation while the others are static ones as background. In such systems, the coupling between the Floquet modes(FMs, originated from the Floquet modulation) and the background modes (BMs, resulted from the background array) will enable peculiar wave dynamics, such as localization-delocalization phase transitions~\cite{PhysRevA.97.023808}, topological protection~\cite{fedorova2019limits}, and the generation of synthetic gauge fields and solitons~\cite{lumer2019light, ivanov2021floquet}. This is schematically shown in Fig.~\ref{fig:schmatic}(b), where the gray symbols indicate the BMs, and the orange symbols represent the FMs. We further note that the corresponding Floquet bands are equidistant due to the periodic time modulation~\cite{he2019floquet, PhysRevApplied.16.014017, PhysRevLett.122.173901}.

Furthermore, when the Floquet system meets non-Hermitian physics, the combination of the two may ignite new ideas, phenomena, and applications, and this is happening in light field manipulation, non-Hermitian band structure, multiple phase transitions~\cite{weidemann2022topological, park2022revealing, PhysRevB.108.014204}, etc. In this paper, we discuss the wave dynamics in a particular arrangement of one-dimensional (1D) lattice with a defect that combines the non-Hermitian and Floquet modulation. The system is a 1D lattice array with fixed and identical potential in all sites (i.e., the background array) except the middle one (i.e., the defect), of which the potential changes periodically in the direction of propagation, as shown in Fig.~\ref{fig:schmatic}(a). In addition, we add gain and loss that satisfy PT-symmetry into the defect. While systems containing only PT-symmetric defects or Floquet modulation defects have been discussed in literature~\cite{fedorova2019limits, PhysRevA.97.023808, Zhou:10, Wang:11}, the wave dynamic behavior in such an arrangement has not yet been discussed. In our system, due to the introduction of PT-symmetric potential into the Floquet modulation, the previous FMs that are originated from the Floquet modulation are further split into gain modes (GMs) and loss modes (LMs), as schematically shown by the change between Figs.~\ref{fig:schmatic}(b) and (c). Now, the coupling is between the BMs and GMs or LMs. We found that, when gain-loss parameter $\Gamma$ has different values, the system will exhibit three phases with different wave behaviors: delocalized wave propagation in phase I, localized propagation with energy oscillation in phase II, and exponential energy boost in the PT-broken phase (phase III), as schematically shown in Fig.~\ref{fig:schmatic}(d). This is different from the behavior of typical PT-symmetric systems, where there are only two phases, i.e., the PT-symmetric and PT-broken phases that are separated by the exceptional point~\cite{feng2017non, liu2024floquet}. This means that the combination of Floquet modulation and PT-symmetry can achieve a new phase with localized and oscillation fields. In addition, it is found that the new localized-oscillation field is jointly controlled by the Floquet modulation and gain-loss parameter $\Gamma$, and the final energy oscillation is the beating of two energy oscillations from PT-symmetric potential and Floquet modulation. Finally, it is interesting to note that, with fixed gain-loss parameter while varied Floquet parameters, the system can also transit from a PT-symmetric phase to a PT-broken phase. These results provide new perspectives on the wave dynamics in systems exhibiting both Floquet modulation and PT-symmetry, and may inspire new applications for light field manipulation.

\section{Wave dynamics in 1D lattice with different defects}
In the proposed 1D lattice array with PT-symmetric Floquet defect, as schematically shown in Fig.~\ref{fig:schmatic}(a), the potential functions of the background $V_b(x,t)$ and the defect $V_d(x,t)$ are written as
\begin{eqnarray}
	V_b(x,t) & = & 4\cos(x)^2,\\ 
	V_d(x,t) & = & 4\left[\cos(x)^2-i \Gamma \sin(2x)\right]+ p \cos(t)^2,\label{potential}
\end{eqnarray}
where, for the defect potential at $|x|\le\pi/2$, $\Gamma$ is the gain-loss parameter that represents the amplitude of the gain and loss, and $p$ is the amplitude of the Floquet modulation with modulation frequency (period) being fixed to $\omega=2~(T=\pi)$. These two parameters control the propagation behavior of the light field inside such a lattice array system. Under the paraxial approximation, the evolution of waves inside the system can be described by the Schrödinger-like equation~\cite{PhysRevLett.100.103904}
\begin{equation}
    i\frac{\partial}{\partial t} \psi(x,t)=-\nabla^2 \psi(x,t)-V(x,t)\psi(x,t),\label{eq:schrodinger}
\end{equation}
In the following subsections, we discuss how the Floquet parameter $p$ and the gain-loss parameter $\Gamma$ independently and collaboratively control the wave dynamics inside the lattice array. We will also discuss the associated wave behaviors and the dual phase transitions.

\subsection{Defect with only Floquet modulation}\label{sec:dFloquet}
Firstly, we analyze the lattice array when there is only Floquet modulation in the defect, i.e., with $\Gamma=0$ and $p\ne0$. In this case, the defect potential in Eq.~(\ref{potential}) is simplified to
\begin{equation}
	V_{F}(x,t) = 4\cos(x)^2+ p \cos(t)^2,\label{eq:VF}
\end{equation}
with the subscript ``$F$'' denotes ``Floquet'' defect, and the corresponding wave dynamic equation in Eq.~(\ref{eq:schrodinger}) can be written as
\begin{align}
    \begin{array}{rcl}
    \displaystyle i\frac{\partial}{\partial t}\psi(x,t) & = & H(x,t)\psi(x,t)\\
     & = & (H_0(x)+H_t(x,t))\psi(x,t),
    \end{array}\label{eq:eigen}
\end{align}
where the total Hamiltonian $H(x,t)$ can be written as the combination of time-independent part $H_0(x)$ and time-dependent part $H_t(x,t)$. Under the Floquet modulation with period $T$, we have $H(x,t)=H(x,t+n T)$, and based on the Floquet theorem, the solution of the system is in the form of 
\begin{equation}
	\psi(x,t)=\phi(x,t)e^{-i\beta t},\label{eq:psiF}
\end{equation}
where $\beta$ is referred as quasi-energy in ${\bm k}$-space and $\phi(x,t)=\phi(x,t+n T)$ is a periodic eigenfunction with the same period as the Hamiltonian. Then, the eigenvalue equation for the quasi-energy can be obtained from Eq.~(\ref{eq:eigen}) as~\cite{PhysRevB.90.195429, PhysRevResearch.3.023211}
\begin{equation}
    \left[H(x,t)-i\frac{\partial}{\partial t}\right]\phi(x,t)=\beta \phi(x,t).\label{eqeig}
\end{equation}
Due to the same periodicity of Hamiltonian $H$ and eigenfunction $\phi$, we can expand them in terms of discrete Fourier series
\begin{equation}
    \begin{array}{rcl}
        H(x,t) & = & \displaystyle \sum_{m} h_m(x)e^{i m\omega t},\\ \\
        \phi(x,t) & = & \displaystyle \sum_{n} u_n(x)e^{i n\omega t},
    \end{array} \label{eqsum}
\end{equation}
with $\omega=2\pi/T=2$, and the coefficients of the harmonics can be calculated by the inverse Fourier transform as  $h_m(x)=(1/T)\int_0^T H(x,t) e^{-i m\omega t}dt$. Substituting them into Eq.~(\ref{eqeig}) we obtain
\begin{equation}
 \sum_{m} \sum_{n} h_m u_n e^{i (m+n)\omega t} + \sum_{n} n \omega u_n e^{i n\omega t}=\beta \sum_{n} u_n e^{i n\omega t}.
   \label{eq:beattotal}
\end{equation}
Replacing $n\rightarrow n-m$ for the first term, and considering the fact that the summations are from $-\infty$ to $+\infty$, the above equation can be simplified to
\begin{equation}
	\sum_{n}\left(\sum_m h_m u_{n-m} + n \omega u_n\right) e^{i n\omega t}=\sum_{n}\beta u_n e^{i n\omega t}.
	\label{eq:beatsingle}
\end{equation}
Therefore, the $n$-th harmonic $e^{i n\omega t}$ couples with the neighboring ones, and we have
\begin{equation}
	\sum_m h_m u_{n-m} + n \omega u_n =\beta u_n.\label{eq:simple}
\end{equation}
Specifically, for the time modulation in the form of $p\cos(t)^2$ with a modulation frequency $\omega=2$ (period $T=\pi$), the coefficients of the Hamiltonian can be obtained as
\begin{equation}
	\begin{array}{lcl}
		h_0 & = & \displaystyle \frac{1}{\pi}\int_{0}^{\pi} (H_0 + p\cos(t)^2)e^{i0t} dt = H_0+p/2, \\ \\
		h_{\pm 1} & = & \displaystyle \frac{1}{\pi} \int_{0}^{\pi} (H_0+p\cos(t)^2) e^{\pm i \omega t} dt = p/4,
	\end{array} \label{fcom}
\end{equation}
and all higher order coefficients being 0. Therefore, for this specific Floquet modulation, Eq.~(\ref{eq:simple}) can be expressed in a matrix form as
\begin{widetext}
\begin{equation}
    \begin{pmatrix}
        ...  & ... & ... & ... & ... & ... & ... \\
        ... & H_0+p/2-2\omega & p/4 & 0 & 0 & 0 & ... \\
        ... & p/4 & H_0+p/2-\omega & p/4 & 0 & 0 & ... \\
        ... & 0 & p/4 & H_0+p/2 & p/4 & 0 & ... \\
        ... & 0 & 0 & p/4 & H_0+p/2+\omega & p/4 & ... \\
        ... & 0 & 0 & 0 & p/4 & H_0+p/2+2\omega & ... \\
        ...  & ... & ... & ... & ... & ... & ... 
    \end{pmatrix}.\left[ \begin{array}{cccc}
    ...\\
    u_{-2}\\
    u_{-1}\\
    u_{0}\\
    u_{1}\\
    u_{2}\\
    ...
\end{array}
\right]=\beta \left[ \begin{array}{cccc}
    ...\\
    u_{-2}\\
    u_{-1}\\
    u_{0}\\
    u_{1}\\
    u_{2}\\
    ...
\end{array}
\right].\label{eq:matrix}
\end{equation}
\end{widetext}

\begin{figure}[!b]
	\centering
	\includegraphics[width=0.6\linewidth]{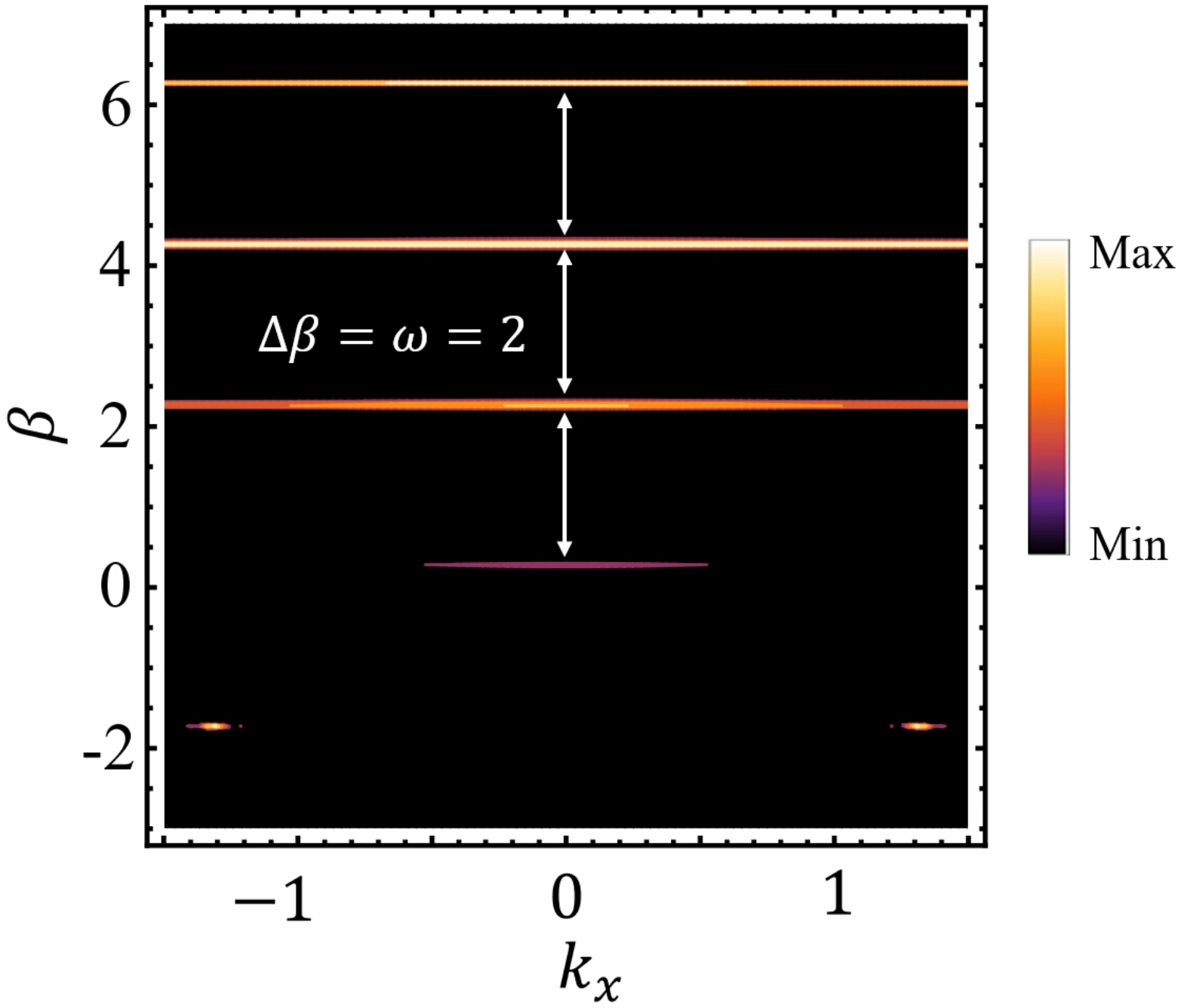}
	\caption{Energy levels for the Floquet defect when the modulation frequency (amplitude) is $\omega=2~(p=4)$.}
	\label{fig:sigsite}
\end{figure}

We can see that, due to the Floquet modulation, the coupling terms between neighboring harmonics are introduced into the Hamiltonian, and the coupling coefficient is $p/4$. In addition, the modulation amplitude $p$ will lift all the energy levels by $p/2$, as shown by the diagonal terms in the Hamiltonian in Eq.~(\ref{eq:matrix}). Such a Hamiltonian will result in equally spaced energy levels in $\bm k$-space, which are called the Floquet bands, and the corresponding FMs will be the superposition of the harmonics. We also note that, the intervals between the Floquet bands are $\omega$. These can be verified from numerical simulation results (all the simulations in this work are performed by the commercial software COMSOL Multiphysics). For example, with modulation frequency $\omega=2$ and amplitude $p=4$, the energy levels obtained by numerical simulation are shown in Fig.~\ref{fig:sigsite}, where the Floquet bands are evenly spaced with interval $\Delta\beta=\omega=2$. 

Now, when this defect is placed in a background periodic array, there will be a background band structure with associated BMs, and the FMs from the Floquet defect will interact with the BMs depending on the positions of the bands. There are two cases describing the interactions: (1) when the Floquet bands overlap with the background bands, the FMs and BMs will couple to each other, and (2) when the background bands fall inside the band gaps of Floquet bands without overlap, the FMs and BMs will be isolated without coupling. This is crucial for understanding the different wave dynamic behaviors in the studied system, and we will discuss in more detail in the following.

\begin{figure}[b]
	\centering
	\includegraphics[width=0.9\linewidth]{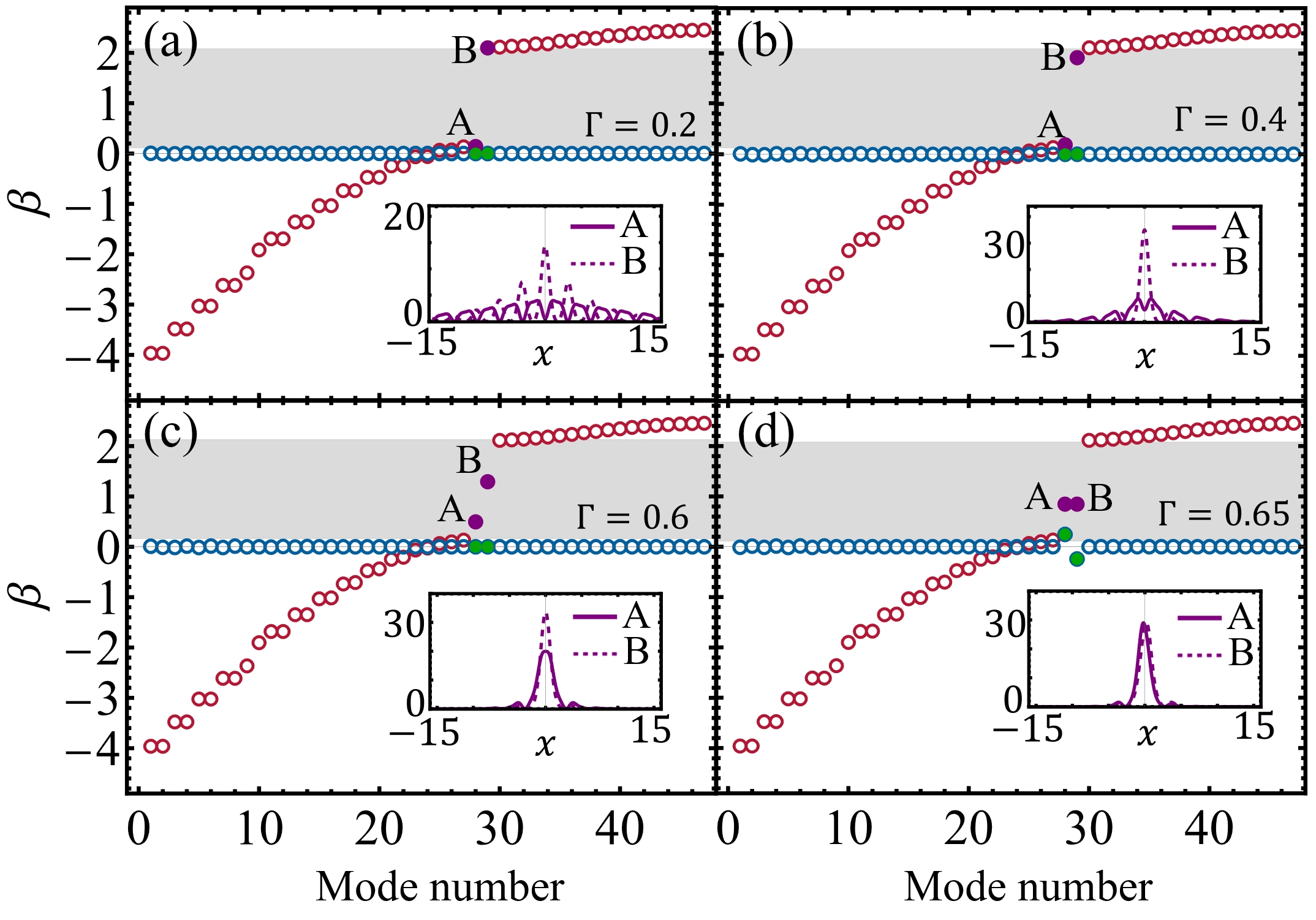}
	\caption{The real and imaginary parts of the eigenvalues for a 19-lattice array when the gain-loss parameter $\Gamma$ of the PT-symmetric defect at the middle is (a) 0.2, (b) 0.4, (c) 0.6, and (d) 0.65, respectively. The red and blue circles (purple and green disks) represent the real and imaginary parts of the background bulk (edge) energy levels. The solid and dashed purple lines in the insets represent the two edge defect modes.}
	\label{fig:staeig}
\end{figure}

\begin{figure*}[t]
	\centering
	\includegraphics[width=0.65\linewidth]{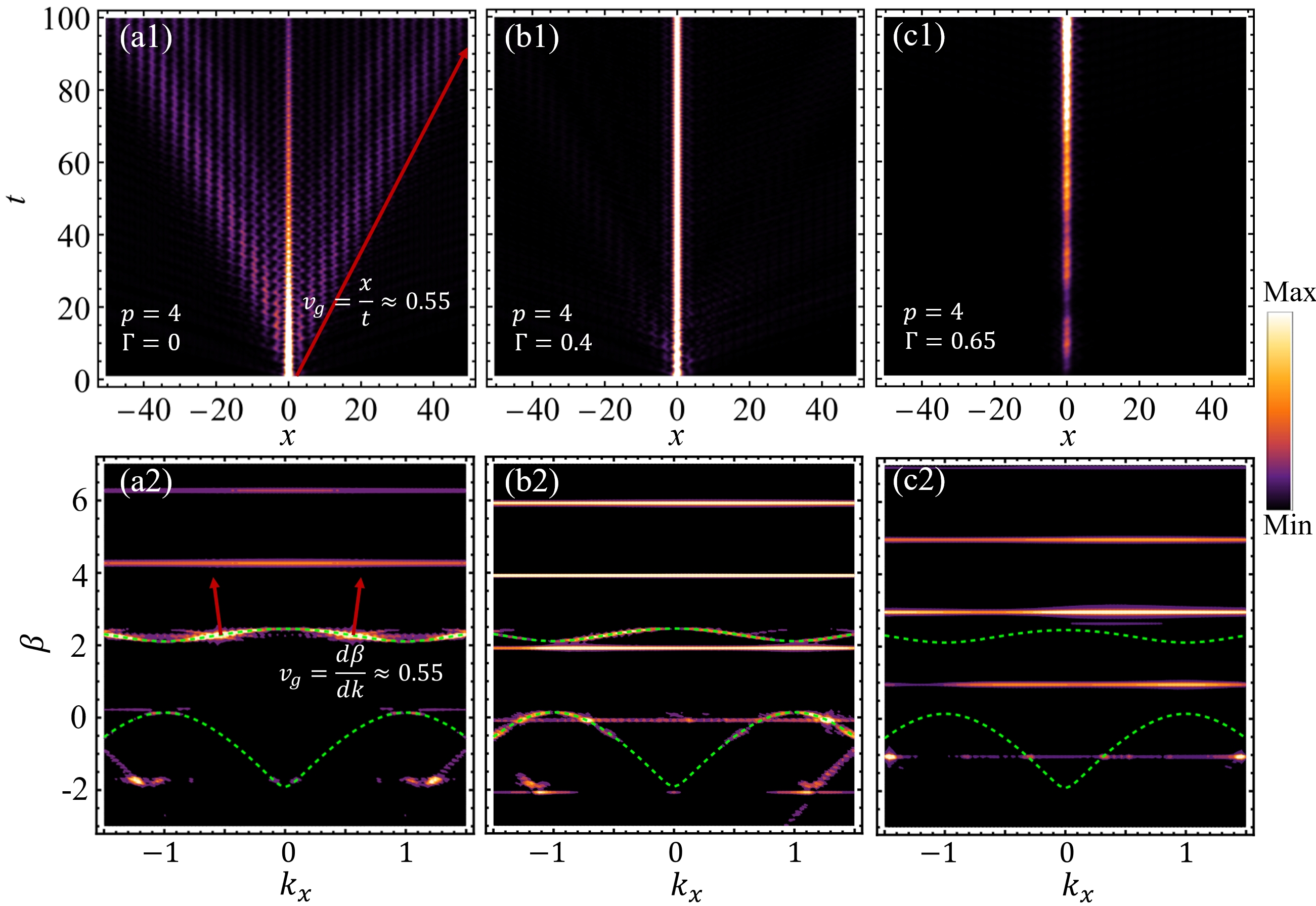}
	\caption{Time domain field distributions (top) and the corresponding energy levels in $\bm k$-space (bottom) for the three phases when the gain-loss parameters $\Gamma$ equals to (a) 0, (b) 0.4, and (c) 0.65, respectively. The green dashed curves are the calculated background bands. The red arrows in (a) represent the diffraction directions of the wave packet.}
	\label{fig:prog1}
\end{figure*}

\subsection{Defect with only PT-symmetric potential}
Secondly, we consider the case when the defect is only with PT-symmetric potential, i.e., with $p=0$ and $\Gamma\ne0$. Then, the defect potential in Eq.~(\ref{potential}) can be rewritten as
\begin{equation}
	V_{PT}(x,t) = 4\left[\cos(x)^2-i \Gamma \sin(2x)\right],\label{eq:VPT}
\end{equation}
with the subscript ``$PT$'' denoting ``PT-symmetric'' defect. We will discuss how the gain-loss parameter $\Gamma$ affects the energy levels and the defect modes. For the PT-symmetric defect with gain-loss, a simple coupled mode matrix can effectively predict and explain the changes in eigenvalues. The Hamiltonian for the defect relating the gain and loss part can be written as
\begin{equation}
	H_{PT}=\begin{pmatrix}
		E_a +i \Gamma_e  & \kappa_e \\
		\kappa_e & E_a -i \Gamma_e 
	\end{pmatrix},\label{eq:HPT}
\end{equation}
where $E_a$ is the real part of the defect site eigenvalue, $\Gamma_e$ is the effective gain-loss coefficient, and $\kappa_e$ represents the effective coupling between the gain and loss parts. The eigenvalues of this matrix are easily obtained as $E_a\pm \sqrt{\kappa_e^2-\Gamma_e^2}$. Obviously, when $\Gamma_e$ increases and across $\kappa_e$, the system will undergo a phase transition from PT-symmetric phase to PT-broken phase, and the eigenvalues are changed from real to complex values. While the background lattice array presents the ``static'' background bulk bands and BMs, the increase of $\Gamma_e$ from the PT-symmetric potential will ``drag'' the edge energy levels that originally belonged to the bulk into the band gap, gradually forming a localized state in the defect. This is common in non-Hermitian systems~\cite{zhen2015spawning, PhysRevLett.131.013802}, and the detailed discussion based on symmetry is also well discussed~\cite{PhysRevLett.120.093901, ge2018non}. 

The formation of the localized states in the lattice array with PT-symmetric defect can be visualized with simulation results. For example, when we have a periodic array of 19 lattices with PT-symmetric defect at the middle, the eigenvalues of the system are plotted in Fig.~\ref{fig:staeig}(a)-(d) for $\Gamma=0.2,0.4,0.6,0.65$, respectively. In this figure, the red and blue circles represent the real and imaginary parts of the bulk eigenvalues, and the purple and green disks mark the real and imaginary parts of the defect edge bands. It can be observed that, with the increase of the gain-loss parameter $\Gamma$, the edge energy levels moves inside the band gap, and finally degenerated when the system goes into the PT-broken phase, as illustrated in Fig.~\ref{fig:staeig}(d). The corresponding edge modes close to the defect are also depicted in the insets of the figures, clearly demonstrating that these are edge modes.

\subsection{Dual phase transitions in the proposed system}
From the above analysis, one can conclude that both Floquet modulation and PT-symmetric potential can cause the movement of defect energy levels. Therefore, the defect with both Floquet modulation and PT-symmetric potential will also move the defect energy levels. As the background bands from the background array are ``static'', the movement of the defect energy levels will result in their overlap or non-overlap with the background bulk bands. Thus, the defect modes will couple or decouple with the BMs, and different wave dynamic behaviors, including delocalized, localized with energy oscillation, and energy boost beam behaviors, will be observed.

In this subsection, we study in detail the wave dynamics and phase transitions in the 1D lattice array with a defect that combining PT-symmetric potential and the Floquet modulation. In this case, depending on the system parameters, there will be three phases, i.e., energy-delocalized phase, energy-localized phase with energy oscillation, and the PT-broken phase with energy boost, accompanying dual phase transitions, i.e., delocalization-localization phase transition and the PT phase transition. For example, when the Floquet modulation is fixed to $p=4$, with the increase of the gain-loss parameter $\Gamma$, three phases with different wave phenomena can be observed: phase I, delocalized wave behavior when $\Gamma<0.37$; phase II, localized wave behavior with energy oscillation when $0.37<\Gamma<0.62$; and phase III (PT-broken phase), energy boost when $\Gamma>0.62$, as shown in the top panels of Fig.~\ref{fig:prog1}, and they are also schematically shown in Fig.~\ref{fig:schmatic}(d). 

These wave behaviors can be well understood by observing the corresponding energy levels in the $\bm k$-space by Fourier transforming the time domain field distributions in Figs.~\ref{fig:prog1}(a1, b1, c1). The calculated energy levels (the color map) are shown in Figs.~\ref{fig:prog1}(a2, b2, c2) respectively, together with the green dashed curves representing the ``static'' background bulk bands. In the delocalized phase (phase I), as shown in Fig.~\ref{fig:prog1}(a) with $\Gamma=0$, it is clear that one of the Floquet bands overlaps with the background bands, resulting in the energy leakage into the background array, exhibiting delocalization behavior. We note that this behavior is different from the discrete diffraction of a typical 1D lattice~\cite{christodoulides2003discretizing}, because only the modes corresponding to the overlapped parts of the bands are excited, rather than the entire band. In addition, the group velocity of the propagating wave packet equals to $v_g=d\beta/dk$, and we can see that it is consistent from the band calculation and the time domain field distribution. On the other hand, in phase II where field is localized inside the defect with energy oscillation, the bands are isolated, as shown in Fig.~\ref{fig:prog1}(b) with $\Gamma=0.4$. This is because the larger gain-loss parameter $\Gamma$ moves the defect bands downward and reach the condition that they are isolated with the main background band. As a result, there is no field coupling, thus no field leakage from the defect modes to the BMs. In addition, when the gain-loss parameter $\Gamma$ further increases, while the bands are still isolated, as shown in Fig.~\ref{fig:prog1}(c), the system will undergo a second phase transition from the PT-symmetric region to the PT-broken phase (phase III), resulting in an energy boost behavior. 

We further note that with the change of the Floquet parameter $p$, the phase I and therefore the first phase transition is not always exist. This is because for certain $p$ values, even when $\Gamma=0$, the shift of the Floquet bands making them do not overlap with the background bulk bands. Therefore, there is no energy leakage from the defect to the background array, and the system behavior directly starts from the localized field at the defect with energy oscillations. This will be discussed in more detail in Sec.~\ref{sec:peffect}.

\begin{figure}[b]
    \centering
    \includegraphics[width=1.02\linewidth]{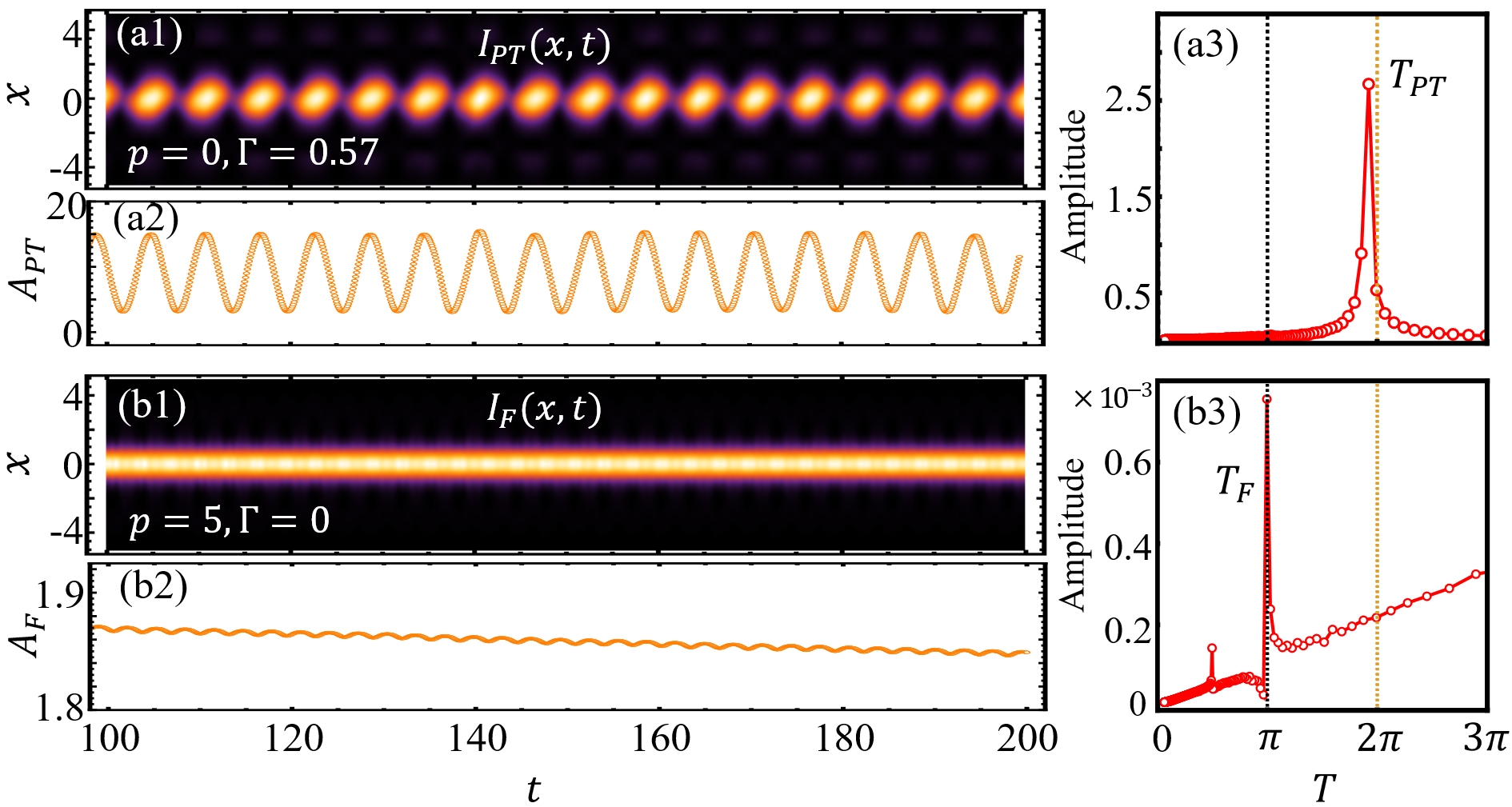}
    \caption{The intensity profile $I(x,t)$ (panel 1), energy amplitude distribution $A(t)$ (panel 2), and the Fourier transform of $A(t)$ showing the oscillation periods (panel 3) for the localized field with energy oscillation when the defect only has (a) PT-symmetric potential with $\Gamma=0.57$ and (b) Floquet modulation with $p=5$.}\label{fig:osc_PT^F}
\end{figure}

\section{Energy oscillations in phase II}
The localized field with energy oscillation in phase II is a non-trivial phenomenon, compared to the behaviors in phases I and III. In phase I, the energy diffuses into the background array. Therefore, the wave packet will evolve according to the eigenstates of the periodic background array, which is trivial. In addition, the energy boost in phase III is also well known as the PT-broken phase. Here, in this section, we analyze the characteristics of the energy oscillation behavior in phase II, which is collaboratively controlled by both Floquet modulation and PT-symmetric potential. This energy oscillation is different from those energy oscillations in the PT-symmetric system or the Floquet defect.

\subsection{Energy oscillation in PT-symmetric defect}\label{sec:osc@PT}
Firstly, we analyze the energy oscillation in a defect with PT-symmetric potential, without Floquet modulation. In this case, there are two modes, i.e. GM and LM with expressions $\psi_{G,L}(x,t)=u_{G,L}(x)e^{i(\beta_{G,L}t+\phi_{G,L})}$, where $\beta_G$ and $\beta_L$ are the two real eigenvalues, as phase II is still in the PT-symmetric region, and $u_{G,L}(x)$ and $\phi_{G,L}$ are the amplitude distributions and phases of the two modes. The fields of these two modes mainly focus inside the defect, as we have shown in Fig.~\ref{fig:staeig}(b) and (c). The overall field distribution is the interference (superposition) of these two modes and the field intensity $I_{PT}(x,t)$ is obtained as:
\begin{equation}
	\begin{array}{l}
	I_{PT}(x,t) = |\psi_G(x,t)+\psi_L(x,t)|^2 \\
	\qquad\qquad\,= u_G^2+u_L^2+2u_Gu_L\cos((\beta_G-\beta_L)t+\Delta\phi).
	\end{array} \label{eq:beat}
\end{equation}
with $\Delta\phi=\phi_G-\phi_L$. One can see that due to the field interference, the intensity will oscillate over time, and the oscillation period is locked to
\begin{equation}
	T_{PT}=2\pi/|\beta_G-\beta_L|.\label{eq:TPT}
\end{equation}
This is verified from simulation results, as shown in Fig.~\ref{fig:osc_PT^F}(a), with gain-loss parameter $\Gamma=0.57$. Figure~\ref{fig:osc_PT^F}(a1) shows the intensity profile $I_{PT}(x,t)$, and Fig.~\ref{fig:osc_PT^F}(a2) shows the amplitude (resembling energy)
\begin{equation}
	\begin{array}{ccl}
		A_{PT}(t) & = & \displaystyle \int I_{PT}(x,t)dx\\
			     & = & \alpha_0 +\alpha \cos((\beta_G-\beta_L)t+\Delta\phi),
	\end{array}
\end{equation}
with $\alpha_0$ and $\alpha$ being coefficients. They both demonstrate the energy oscillation over time. Furthermore, by taking Fourier Transform of the amplitude $A_{PT}(t)$, we can extract the oscillation period $T_{PT}$, which is close to $2\pi$, as shown in Fig.~\ref{fig:osc_PT^F}(a3). We note that the period $T_{PT}$ changes with the change of gain-loss parameter $\Gamma$. Specifically, when $\Gamma$ increases from 0 to the critical value that corresponds to the exceptional point, the period $T_{PT}$ will increase from a finite value to infinity. This is because with the increase of $\Gamma$, the difference between the two real eigenvalues $|\beta_G-\beta_L|$ will decrease from a finite value and reaches to 0 at the exceptional point. We further note that the amplitude of the energy oscillation at period $T_{PT}$, as shown by the peak value in Fig.~\ref{fig:osc_PT^F}(a3), will also increase with the increase of $\Gamma$. This is resulted from the effects of the larger gain and loss, together with the larger period $T_{PT}$. As a result, the energy will amplify (decay) more quickly, and with longer time, to a larger (smaller) value at the gain (loss) part.

\begin{figure}[b]
	\centering
	\includegraphics[width=1.02\linewidth]{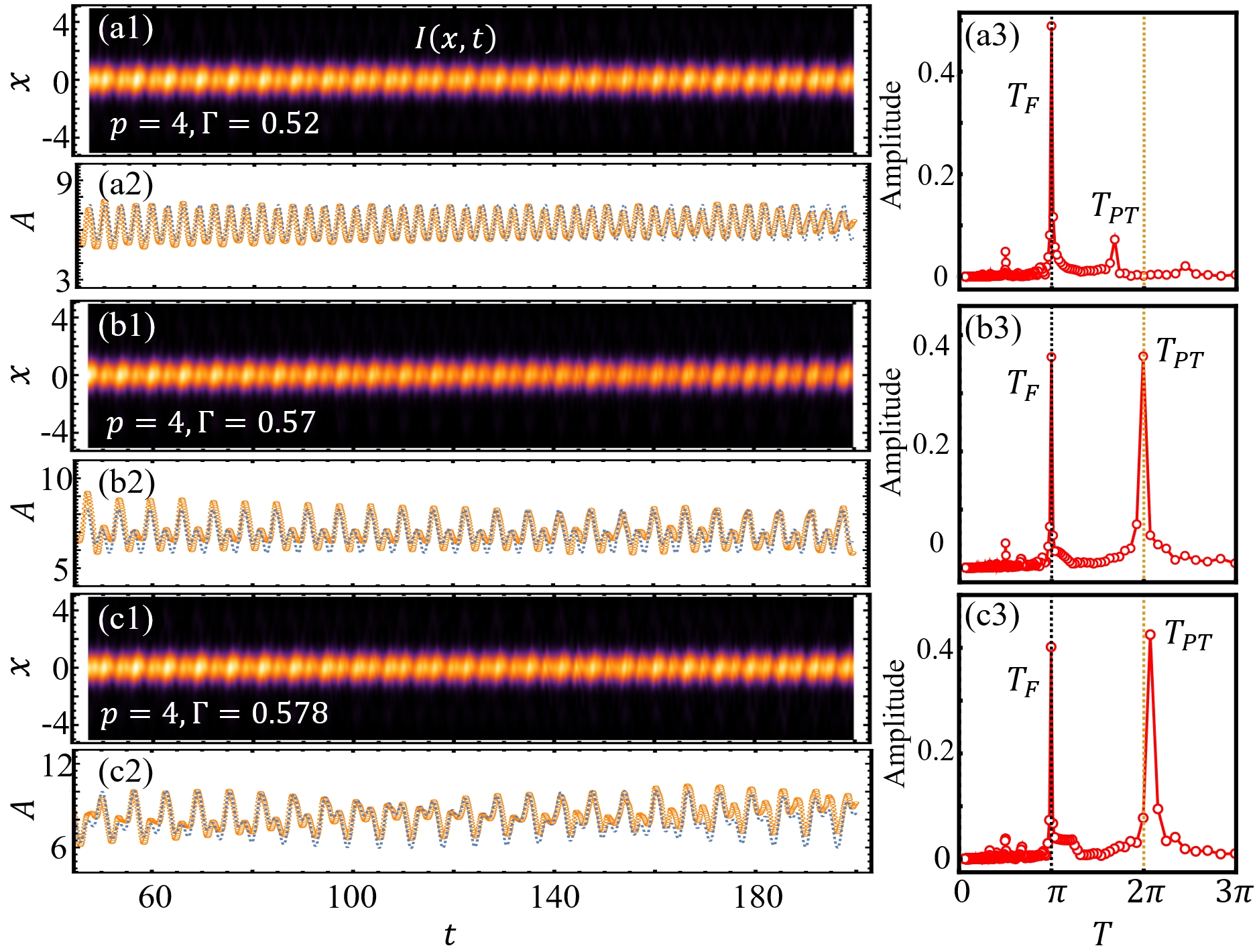}
	\caption{The intensity profile $I(x,t)$ (panel 1), energy amplitude distribution $A(t)$ (panel 2), and the Fourier transform of $A(t)$ showing the oscillation periods (panel 3) for the localized field with energy oscillation when the defect has modulation amplitude $p=4$ and gain-loss parameter $\Gamma$ equals to (a) 0.52, (b) 0.57, and (c) 0.578, respectively. The light blue dashed lines in (a2), (b2), and (c2) are the beating energy functions with the extracted periods $T_F=\pi$ and $T_{PT}$.}
	\label{fig:osc_total}
\end{figure}

\subsection{Energy oscillation in Floquet defect}
Secondly, if there is only Floquet modulation, the superposition of the Floquet modes will also result in energy oscillations. In this case, based on the expressions of the Floquet modes from Eqs.~(\ref{eq:psiF}) and (\ref{eqsum}), the intensity of the total field can be written, similar to Eq.~(\ref{eq:beat}), as
\begin{equation}
	\begin{array}{ccl}
		I_F(x,t) & = & \left|\displaystyle\sum_{n=-\infty}^{\infty} u_n(x) e^{i(\beta_0+n\omega)t}\right|^2\\
				 & = & \displaystyle\sum_{n=-\infty}^{\infty} u_n(x)^2 + \sum_{n=1}^{\infty} v_n(x) \cos(n\omega t),
	\end{array}\label{eq:IF}
\end{equation}
where $\beta_0$ is the fundamental eigenvalue (energy level), $u_n(x)$ is the amplitude distribution of the $n$-th harmonic, and $v_n(x)=\sum_\mu\sum_\nu u_\mu(x)u_\nu(x)$ being a linear combination of $u_\mu(x)u_\nu(x)$ for different mode numbers $\mu$ and $\nu$. From Eq.~(\ref{eq:IF}) we can see that the frequencies of the energy oscillation due to Floquet modulation are multiples of the modulation frequency $\omega$. Therefore, the oscillation periods will be 
\begin{equation}
	T_F=2\pi/n\omega.\label{eq:TF}
\end{equation}
For example, if the frequency of the Floquet modulation is $\omega=2$ as described in Eq.~(\ref{eq:VF}), the periods of the energy oscillation will be $\pi/n$. This is verified with simulation results, as shown in Fig.~\ref{fig:osc_PT^F}(b), where the fields are localized inside the Floquet defect when $p=5$, and the Fourier transform of the $A_F(t)$ clearly demonstrating oscillation periods $\pi$ and $\pi/2$.

\subsection{Energy oscillation in PT-symmetric Floquet defect}
Finally, if the defect has both PT-symmetric potential and Floquet modulation, we found that the final energy oscillation is the beating of the energy oscillations from PT-symmetric potential and Floquet modulation, rather than resulting from the field interference between the Floquet modes and the GM/LM from PT-symmetric potential. In other words, the total intensity profile can be written as
\begin{equation}
	I(x,t)=c_{PT} I_{PT}(x,t)+c_F I_{F}(x,t),\label{eq:Itotal}
\end{equation} 
with $c_{PT}$ and $c_F$ being the coefficients. Therefore, the periods of the energy oscillation include both $T_{PT}$ and $T_F$. For example, when the defect potential in Eq.~(\ref{potential}) is with $p=4$ and $0.37<\Gamma<0.62$ so that the system is in phase II, Fig.~\ref{fig:osc_total} shows the intensity profiles and the extracted periods for $\Gamma$ equals to (a) 0.52, (b) 0.57, and (c) 0.578, respectively.

\begin{figure}[b]
    \centering
    \includegraphics[width=0.5\linewidth]{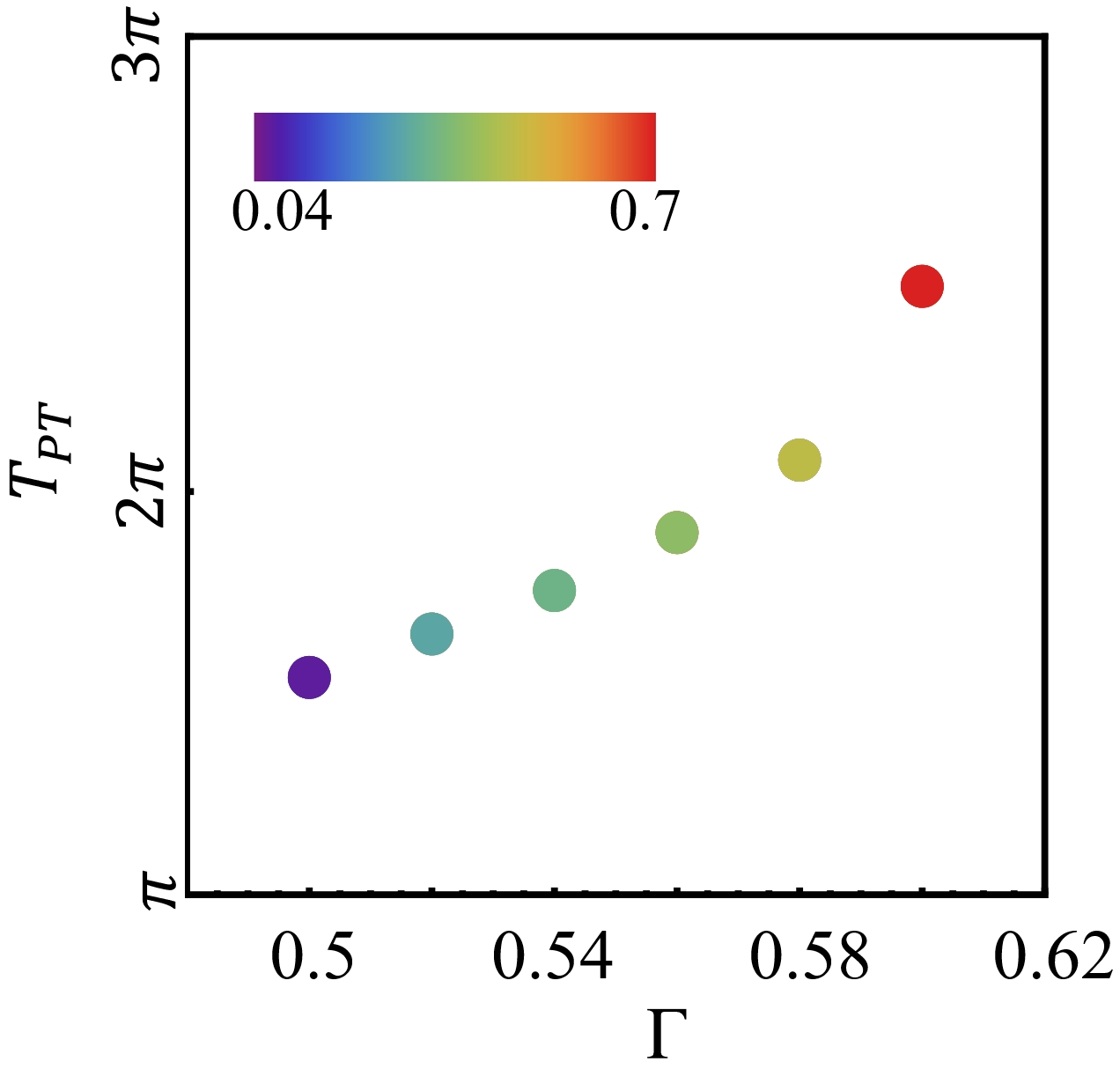}
    \caption{The oscillation period $T_{PT}$ (left axis) and the corresponding oscillation amplitude (color map) versus the gain-loss parameters $\Gamma$ of the PT-symmetric Floquet defect when the amplitude of the Floquet modulation is $p=4$.}
    \label{fig:TPTvGamma}
\end{figure}

From Fig.~\ref{fig:osc_total} one can find that energy exhibits different oscillatory behaviors for different $\Gamma$, and the oscillation periods are identified from the Fourier transform of $A(t)$. On one hand, one can find that in all three cases, the oscillation periods of $\pi$ and $\pi/2$ always exist, and they correspond to the $T_F$ according to the Floquet modulation. In addition, according to panel (3) in Fig.~\ref{fig:osc_total}, the component of $T_F=\pi$ oscillation has high amplitude (around 0.4) while the amplitude of the $T_F=\pi/2$ oscillation is relatively low (around 0.04). On the other hand, the oscillation period of $T_{PT}$ is also observed, and $T_{PT}$ increases with the increase of $\Gamma$. Moreover, the oscillation amplitude of the $T_{PT}$ component also increases with the increase of $\Gamma$. These behaviors are also summarized in Fig.~\ref{fig:TPTvGamma}, and they agree well with the conclusions in Sec.~\ref{sec:osc@PT}.

From these results we can conclude that when the defect is a PT-symmetric Floquet defect, the final energy oscillation is the beating of the energy oscillations from PT-symmetric potential and the Floquet modulation, as described by Eq.~(\ref{eq:Itotal}). This can be observed more clearly from the amplitude distribution $A(t)$ in Fig.~\ref{fig:osc_total}(b2) and (c2). In Fig.~\ref{fig:osc_total}(b2), the oscillation period $T_{PT}$ is very close to the twice of $T_F$, thus the $A(t)$ shows both the periods. While in Fig.~\ref{fig:osc_total}(c2), the oscillation period $T_{PT}$ is slightly larger than $2T_F$, as a result, the $A(t)$ shows the beating of the two energy oscillations. To show the energy beating, the blue dashed lines are plotted according to Eq.~(\ref{eq:Itotal}) with the extracted periods, and they agree with the oscillation very well.

\section{Effect of other Floquet parameters}
\begin{figure}[b]
	\centering
	\includegraphics[width=1.02\linewidth]{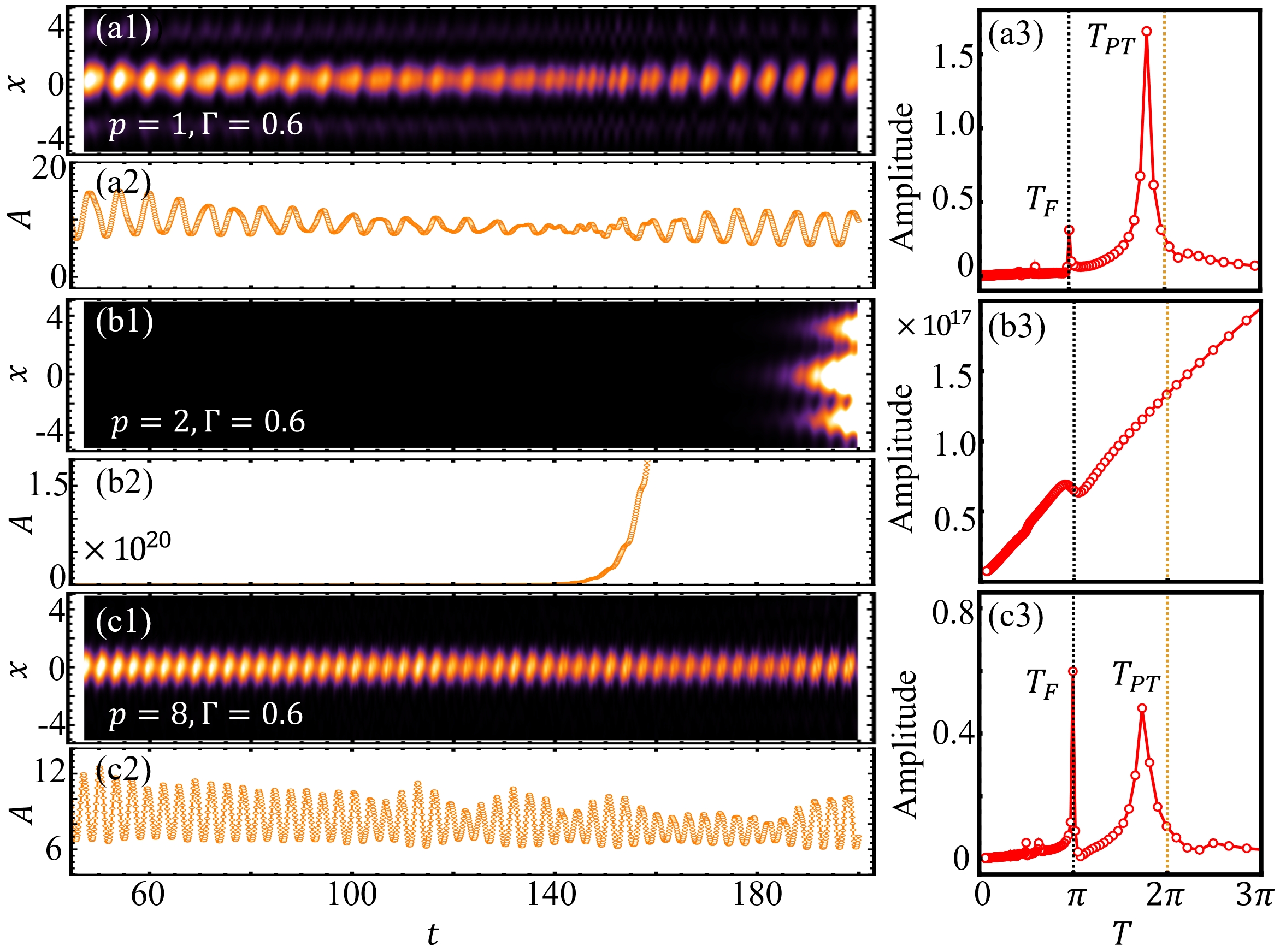}
	\caption{The intensity profile $I(x,t)$ (panel 1), energy amplitude distribution $A(t)$ (panel 2), and the Fourier transform of $A(t)$ showing the oscillation periods (panel 3) for the wave behavior in the studied lattice array when the defect has gain-loss parameter $\Gamma=0.6$ and modulation amplitude $p$ equals to (a) 1, (b) 2, and (c) 8, respectively.}
	\label{fig:progPTRe}
\end{figure}

From the discussions on the energy levels of the Floquet defect from Eq.~(\ref{eq:matrix}) in Sec.~\ref{sec:dFloquet}, we concluded that the Floquet energy levels will shift up or down by changing the modulation amplitude $p$, and the intervals between these Floquet bands will also change by changing the modulation frequency $\omega$ as the intervals are $\Delta\beta=\omega$. Hence, under the change of $p$ or $\omega$, the shift and movement of the Floquet bands will certainly change the wave dynamics in the studied lattice array, i.e., change the phase of the system. In this section, we study in detail the effects of $p$ and $\omega$ on the wave dynamics and the phase transitions.

\subsection{Effect of Floquet modulation amplitude}\label{sec:peffect}
First of all, when there is only Floquet modulation in the defect, the change of modulation amplitude $p$ can already transit the studied system from phase I to phase II. This can be observed by comparing Fig.~\ref{fig:prog1}(a) and Fig.~\ref{fig:osc_PT^F}(b). While both of them have $\Gamma=0$, the change from $p=4$ in Fig.~\ref{fig:prog1}(a) to $p=5$ in Fig.~\ref{fig:osc_PT^F}(b) make the wave behavior from delocalized phase to localized phase. This is because the increase of $p$ shifts the Floquet bands upward, and when $p=5$, there is no overlap between the Floquet bands and the background bulk band, forcing the fields localized inside the defect. From this point of view, the phase I and thus the first phase transition do not exist by properly selecting the modulation amplitude $p$. However, we note that the second PT phase transition always exists.

\begin{figure}[b]
	\centering
	\includegraphics[width=0.7\linewidth]{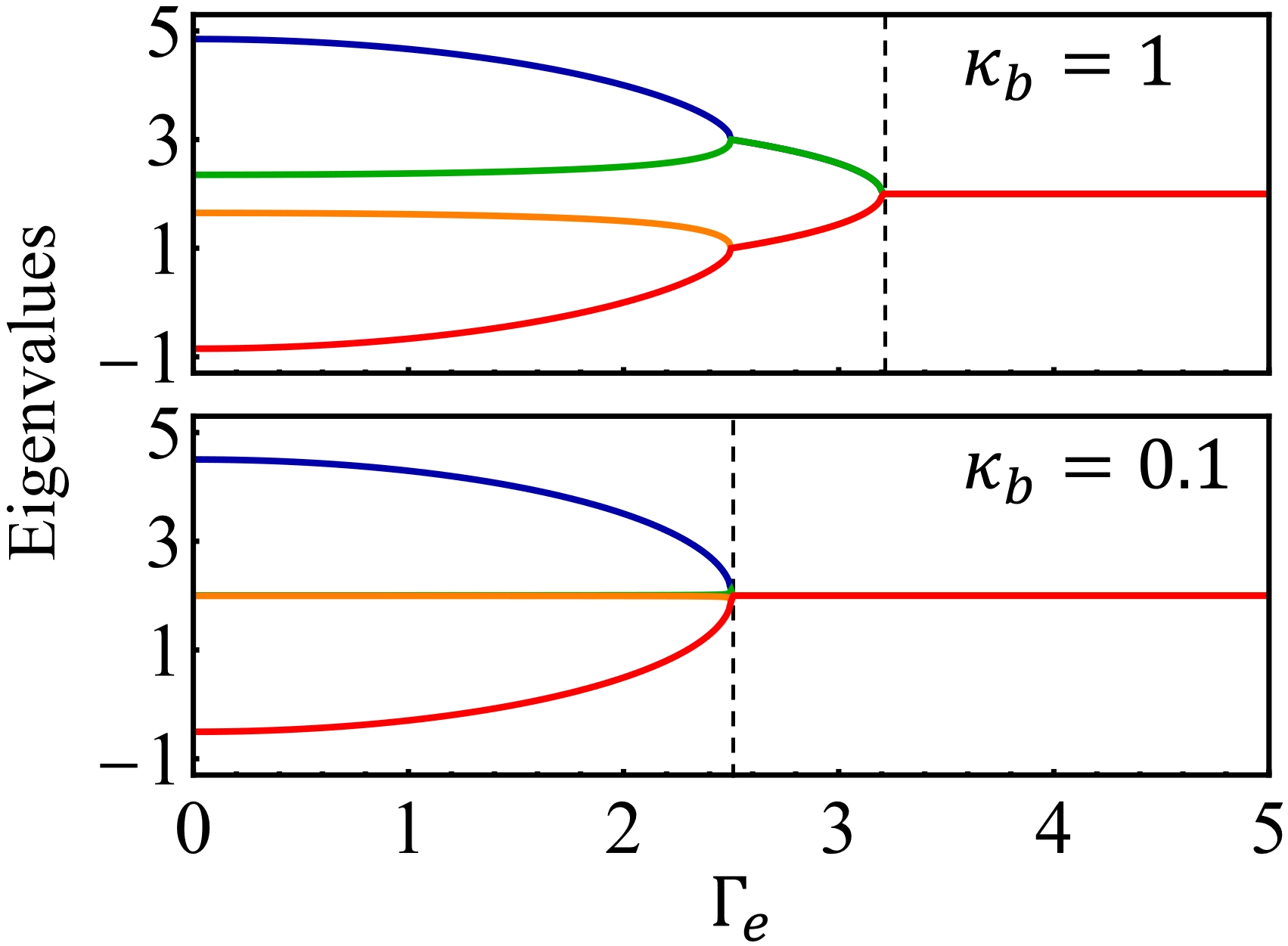}
	\caption{The eigenvalues (real part) of $H_c$ versus the change of effective gain-loss coefficient $\Gamma_e$ for $\kappa_b=1$ (top) and $\kappa_b=0.1$ (bottom), with $E_a=E_b=2$ and $\kappa_e=2.5$. The vertical black dashed lines show the PT-symmetry breaking point.}
	\label{fig:cmeig}
\end{figure}

\begin{figure*}
	\includegraphics[width=0.9\textwidth]{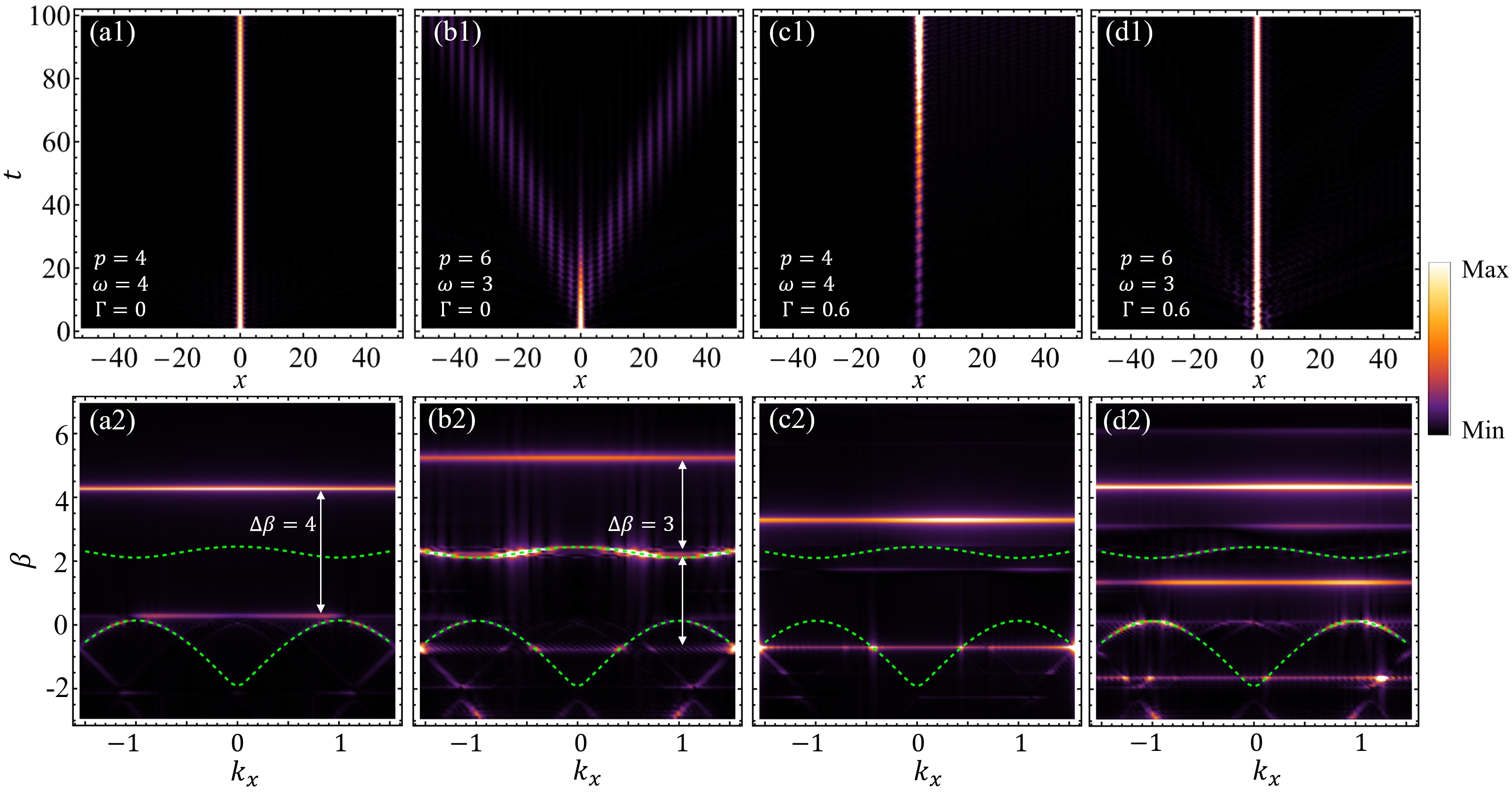}
	\caption{Time domain field distributions (top) and the corresponding energy levels in $\bm k$-space (bottom) for different Floquet and PT-symmetric parameters, as listed in the left bottom corner of the top panels.}\label{fig:omega}
\end{figure*}

On the other hand, when the PT-symmetric potential is also exist in the defect, the phase transition points of the dual phase transitions will also be affected. For example, when we fix the gain-loss parameter $\Gamma=0.6$ and vary the Floquet modulation parameter $p$, Fig.~\ref{fig:progPTRe} depicts the wave dynamic behaviors for $p=1,~2,~8$ respectively. It is clear that, when modulation amplitude $p$ changes from 1 to 2, and to 8, the system is in the phase II, phase III, and phase II again, as shown in Figs~\ref{fig:progPTRe}(a), (b) and (c), respectively. 

The above behavior can be understood as that, for different modulation amplitude $p$, the PT phase transition point corresponds to different $\Gamma$ values. It can be explained by a mode-coupling model between the PT-symmetric Floquet defect and the background lattices. While the PT-symmetric potential of the defect can be modeled as Eq.~(\ref{eq:HPT}), the effect of the Floquet modulation amplitude $p$ will change the coupling $\kappa_b$ between the defect and the background lattices. This is because, the periodic Floquet bands will move up with the increase of $p$, making the Floquet modes periodically couple to the BMs. Thus, $\kappa_b$ will also change periodically with the increase of $p$. Therefore, the Hamiltonian describing this effect can be written as
\begin{equation}
	H_c=
	\begin{pmatrix}
		E_b  & \kappa_b & 0 & 0 \\
		\kappa_b & E_a+i \Gamma_e & \kappa_e & 0 \\
		0 & \kappa_e & E_a-i\Gamma_e & \kappa_b \\
		0 & 0 & \kappa_b & E_b
	\end{pmatrix},
\end{equation}
with $E_b$ being the real part of the eigenvalue of the background lattice, which equals to $E_a$. For this Hamiltonian, Fig.~\ref{fig:cmeig} shows the change of its eigenvalues (real part) with the change of effective gain-loss parameter $\Gamma_e$ for two couplings $\kappa_e=1$ and $0.1$. As we can see, the PT phase transition point (the second phase transition) corresponds to larger $\Gamma_e$ for larger $\kappa_b$. For the field behaviors in Fig.~\ref{fig:progPTRe}, we can explain with the mapping that $\Gamma=0.6$ in Fig.~\ref{fig:progPTRe} relates to $\Gamma_e=3$ in the model, and $p=1,~8$ ($2$) correspond to the strong (weak) coupling $\kappa_b$, resulting in the wave behaviors in phase II (phase III). Reversely thinking, if we change the background lattice array, the coupling $\kappa_b$ will also be changed. Thus it is worth noting that controlling the periodic array can also control the dynamics of wave packets in the studied lattice system. 

\subsection{Effect of Floquet modulation frequency}
With a fixed modulation amplitude $p$, the change of the Floquet modulation frequency $\omega$ will also transit the wave dynamics between different phases, as the Floquet bands are also moved according to the change of intervals $\Delta\beta=\omega$. To study the effect of $\omega$ on the wave behavior, we rewrite the defect potential in Eq.~(\ref{potential}) into a general form
\begin{equation}
	V_d(x,t) = 4\left[\cos(x)^2-i \Gamma \sin(2x)\right]+ p \cos\left(\frac{\omega}{2}t\right)^2.\label{eq:Vdfull}
\end{equation}
We note that $\omega=2$ is used for all the previous cases. First of all, when there is only Floquet modulation in the defect, i.e., with $\Gamma=0$, Figs.~\ref{fig:omega}(a) and (b) show the time domain field distributions (top) and the corresponding energy levels in the $\bm k$-space (bottom) for different Floquet parameters: $p=4$ and $\omega=4$ in Figs.~\ref{fig:omega}(a), and $p=6$ and $\omega=3$ in Figs.~\ref{fig:omega}(b). Clearly, from the field distributions (top panels) we can see that the system is in phase II and phase I, respectively. These wave behaviors are indeed controlled by the positions of the Floquet bands, for which the intervals are $\Delta\beta=\omega$ and their positions are also affected by $p$, as clearly shown by the corresponding energy levels in the $\bm k$-space (bottom panels). Then, with the introduction of PT-symmetric potential with $\Gamma=0.6$, the system will transit into phase III (phase II) by crossing the second (first) phase transition, as clearly shown in Figs.~\ref{fig:omega}(c) and (d). 

In summary, in our studied 1D lattice array with defect, the phases and the corresponding dual phase transitions are collaborative controlled by the Floquet modulation and PT-symmetric potential in the defect. The Floquet parameters $p$ and $\omega$, and the gain-loss parameter $\Gamma$, determine the energy levels of the defect. And depending on the relative positions of the defect energy levels and background bands, different wave dynamic behaviors in the three phases are achieved with the corresponding phase transitions. 

\section{Conclusion}
In this work, we study in detail the wave dynamics in a simple 1D lattice array with a PT-symmetric Floquet defect. Under the change of the control parameters, i.e., the gain-loss parameter $\Gamma$, the Floquet modulation parameters $p$-amplitude and $\omega$-frequency, the system undergoes dual phase transitions from phase I (energy-delocalized phase), to phase II (localized phase with energy oscillations), and finally phase III (PT-symmetry broken phase with energy boost). We find that the different phases are controlled by the relative position of the background bulk bands, which are ``static'' due to the background lattice array, and the defect energy levels, which are determined by the control parameters. When the defect bands overlap with the bulk bands, the defect modes can couple to the BMs, thus resulting in delocalized wave propagation. Whereas when they do not overlap with each other, the fields will be localized inside the defect, resulting in the phase II or phase III depending on their relative positions. In addition, we also study the energy oscillations in the phase II. We find that the final energy oscillation is the beating of the two energy oscillations, i.e., the energy oscillation from the PT-symmetric potential with period $T_{PT}$ and the energy oscillation from the Floquet modulation with periods $T_F=2\pi/n\omega$. This is different from those usual energy oscillations due to the field interference or mode superposition. These results extends the the study of non-Hermitian Floquet systems, and may ignite new ideas for light field manipulation with the collaboration of Floquet modulation and PT-symmetry.

\begin{acknowledgments}
This work was mainly supported by the National Natural Science Foundation of China under Grant No. 12274339. The authors also acknowledge financial supports from Shaanxi Province and Xi’an Jiaotong University.
\end{acknowledgments}

\appendix



\bibliography{references}

\begin{thebibliography}{68}%
\makeatletter
\providecommand \@ifxundefined [1]{%
 \@ifx{#1\undefined}
}%
\providecommand \@ifnum [1]{%
 \ifnum #1\expandafter \@firstoftwo
 \else \expandafter \@secondoftwo
 \fi
}%
\providecommand \@ifx [1]{%
 \ifx #1\expandafter \@firstoftwo
 \else \expandafter \@secondoftwo
 \fi
}%
\providecommand \natexlab [1]{#1}%
\providecommand \enquote  [1]{``#1''}%
\providecommand \bibnamefont  [1]{#1}%
\providecommand \bibfnamefont [1]{#1}%
\providecommand \citenamefont [1]{#1}%
\providecommand \href@noop [0]{\@secondoftwo}%
\providecommand \href [0]{\begingroup \@sanitize@url \@href}%
\providecommand \@href[1]{\@@startlink{#1}\@@href}%
\providecommand \@@href[1]{\endgroup#1\@@endlink}%
\providecommand \@sanitize@url [0]{\catcode `\\12\catcode `\$12\catcode
  `\&12\catcode `\#12\catcode `\^12\catcode `\_12\catcode `\%12\relax}%
\providecommand \@@startlink[1]{}%
\providecommand \@@endlink[0]{}%
\providecommand \url  [0]{\begingroup\@sanitize@url \@url }%
\providecommand \@url [1]{\endgroup\@href {#1}{\urlprefix }}%
\providecommand \urlprefix  [0]{URL }%
\providecommand \Eprint [0]{\href }%
\providecommand \doibase [0]{https://doi.org/}%
\providecommand \selectlanguage [0]{\@gobble}%
\providecommand \bibinfo  [0]{\@secondoftwo}%
\providecommand \bibfield  [0]{\@secondoftwo}%
\providecommand \translation [1]{[#1]}%
\providecommand \BibitemOpen [0]{}%
\providecommand \bibitemStop [0]{}%
\providecommand \bibitemNoStop [0]{.\EOS\space}%
\providecommand \EOS [0]{\spacefactor3000\relax}%
\providecommand \BibitemShut  [1]{\csname bibitem#1\endcsname}%
\let\auto@bib@innerbib\@empty
\bibitem [{\citenamefont {Sondhi}\ \emph {et~al.}(1997)\citenamefont {Sondhi},
  \citenamefont {Girvin}, \citenamefont {Carini},\ and\ \citenamefont
  {Shahar}}]{RevModPhys.69.315}%
  \BibitemOpen
  \bibfield  {author} {\bibinfo {author} {\bibfnamefont {S.~L.}\ \bibnamefont
  {Sondhi}}, \bibinfo {author} {\bibfnamefont {S.~M.}\ \bibnamefont {Girvin}},
  \bibinfo {author} {\bibfnamefont {J.~P.}\ \bibnamefont {Carini}},\ and\
  \bibinfo {author} {\bibfnamefont {D.}~\bibnamefont {Shahar}},\ }\bibfield
  {title} {\bibinfo {title} {Continuous quantum phase transitions},\ }\href
  {https://doi.org/10.1103/RevModPhys.69.315} {\bibfield  {journal} {\bibinfo
  {journal} {Rev. Mod. Phys.}\ }\textbf {\bibinfo {volume} {69}},\ \bibinfo
  {pages} {315} (\bibinfo {year} {1997})}\BibitemShut {NoStop}%
\bibitem [{\citenamefont {Zurek}\ \emph {et~al.}(2005)\citenamefont {Zurek},
  \citenamefont {Dorner},\ and\ \citenamefont
  {Zoller}}]{PhysRevLett.95.105701}%
  \BibitemOpen
  \bibfield  {author} {\bibinfo {author} {\bibfnamefont {W.~H.}\ \bibnamefont
  {Zurek}}, \bibinfo {author} {\bibfnamefont {U.}~\bibnamefont {Dorner}},\ and\
  \bibinfo {author} {\bibfnamefont {P.}~\bibnamefont {Zoller}},\ }\bibfield
  {title} {\bibinfo {title} {Dynamics of a quantum phase transition},\ }\href
  {https://doi.org/10.1103/PhysRevLett.95.105701} {\bibfield  {journal}
  {\bibinfo  {journal} {Phys. Rev. Lett.}\ }\textbf {\bibinfo {volume} {95}},\
  \bibinfo {pages} {105701} (\bibinfo {year} {2005})}\BibitemShut {NoStop}%
\bibitem [{\citenamefont {Sebasti\'an}\ \emph {et~al.}(2020)\citenamefont
  {Sebasti\'an}, \citenamefont {Cmok}, \citenamefont {Mandle}, \citenamefont
  {de~la Fuente}, \citenamefont {Dreven\ifmmode \check{s}\else~\v{s}\fi{}ek
  Olenik}, \citenamefont {\ifmmode \check{C}\else
  \v{C}\fi{}opi\ifmmode~\check{c}\else \v{c}\fi{}},\ and\ \citenamefont
  {Mertelj}}]{PhysRevLett.124.037801}%
  \BibitemOpen
  \bibfield  {author} {\bibinfo {author} {\bibfnamefont {N.}~\bibnamefont
  {Sebasti\'an}}, \bibinfo {author} {\bibfnamefont {L.}~\bibnamefont {Cmok}},
  \bibinfo {author} {\bibfnamefont {R.~J.}\ \bibnamefont {Mandle}}, \bibinfo
  {author} {\bibfnamefont {M.~R.}\ \bibnamefont {de~la Fuente}}, \bibinfo
  {author} {\bibfnamefont {I.}~\bibnamefont {Dreven\ifmmode
  \check{s}\else~\v{s}\fi{}ek Olenik}}, \bibinfo {author} {\bibfnamefont
  {M.}~\bibnamefont {\ifmmode \check{C}\else
  \v{C}\fi{}opi\ifmmode~\check{c}\else \v{c}\fi{}}},\ and\ \bibinfo {author}
  {\bibfnamefont {A.}~\bibnamefont {Mertelj}},\ }\bibfield  {title} {\bibinfo
  {title} {Ferroelectric-ferroelastic phase transition in a nematic liquid
  crystal},\ }\href {https://doi.org/10.1103/PhysRevLett.124.037801} {\bibfield
   {journal} {\bibinfo  {journal} {Phys. Rev. Lett.}\ }\textbf {\bibinfo
  {volume} {124}},\ \bibinfo {pages} {037801} (\bibinfo {year}
  {2020})}\BibitemShut {NoStop}%
\bibitem [{\citenamefont {Xu}\ \emph {et~al.}(2022{\natexlab{a}})\citenamefont
  {Xu}, \citenamefont {Wu}, \citenamefont {Ye}, \citenamefont {Luo},
  \citenamefont {Jian},\ and\ \citenamefont {Xu}}]{PhysRevX.12.021067}%
  \BibitemOpen
  \bibfield  {author} {\bibinfo {author} {\bibfnamefont {Y.}~\bibnamefont
  {Xu}}, \bibinfo {author} {\bibfnamefont {X.-C.}\ \bibnamefont {Wu}}, \bibinfo
  {author} {\bibfnamefont {M.}~\bibnamefont {Ye}}, \bibinfo {author}
  {\bibfnamefont {Z.-X.}\ \bibnamefont {Luo}}, \bibinfo {author} {\bibfnamefont
  {C.-M.}\ \bibnamefont {Jian}},\ and\ \bibinfo {author} {\bibfnamefont
  {C.}~\bibnamefont {Xu}},\ }\bibfield  {title} {\bibinfo {title}
  {Interaction-driven metal-insulator transition with charge
  fractionalization},\ }\href {https://doi.org/10.1103/PhysRevX.12.021067}
  {\bibfield  {journal} {\bibinfo  {journal} {Phys. Rev. X}\ }\textbf {\bibinfo
  {volume} {12}},\ \bibinfo {pages} {021067} (\bibinfo {year}
  {2022}{\natexlab{a}})}\BibitemShut {NoStop}%
\bibitem [{\citenamefont {Zou}\ and\ \citenamefont
  {Chowdhury}(2020)}]{PhysRevResearch.2.032071}%
  \BibitemOpen
  \bibfield  {author} {\bibinfo {author} {\bibfnamefont {L.}~\bibnamefont
  {Zou}}\ and\ \bibinfo {author} {\bibfnamefont {D.}~\bibnamefont
  {Chowdhury}},\ }\bibfield  {title} {\bibinfo {title} {Deconfined
  metal-insulator transitions in quantum hall bilayers},\ }\href
  {https://doi.org/10.1103/PhysRevResearch.2.032071} {\bibfield  {journal}
  {\bibinfo  {journal} {Phys. Rev. Res.}\ }\textbf {\bibinfo {volume} {2}},\
  \bibinfo {pages} {032071} (\bibinfo {year} {2020})}\BibitemShut {NoStop}%
\bibitem [{\citenamefont {Roushan}\ \emph {et~al.}(2014)\citenamefont
  {Roushan}, \citenamefont {Neill}, \citenamefont {Chen}, \citenamefont
  {Kolodrubetz}, \citenamefont {Quintana}, \citenamefont {Leung}, \citenamefont
  {Fang}, \citenamefont {Barends}, \citenamefont {Campbell}, \citenamefont
  {Chen} \emph {et~al.}}]{roushan2014observation}%
  \BibitemOpen
  \bibfield  {author} {\bibinfo {author} {\bibfnamefont {P.}~\bibnamefont
  {Roushan}}, \bibinfo {author} {\bibfnamefont {C.}~\bibnamefont {Neill}},
  \bibinfo {author} {\bibfnamefont {Y.}~\bibnamefont {Chen}}, \bibinfo {author}
  {\bibfnamefont {M.}~\bibnamefont {Kolodrubetz}}, \bibinfo {author}
  {\bibfnamefont {C.}~\bibnamefont {Quintana}}, \bibinfo {author}
  {\bibfnamefont {N.}~\bibnamefont {Leung}}, \bibinfo {author} {\bibfnamefont
  {M.}~\bibnamefont {Fang}}, \bibinfo {author} {\bibfnamefont {R.}~\bibnamefont
  {Barends}}, \bibinfo {author} {\bibfnamefont {B.}~\bibnamefont {Campbell}},
  \bibinfo {author} {\bibfnamefont {Z.}~\bibnamefont {Chen}}, \emph {et~al.},\
  }\bibfield  {title} {\bibinfo {title} {Observation of topological transitions
  in interacting quantum circuits},\ }\href@noop {} {\bibfield  {journal}
  {\bibinfo  {journal} {Nature}\ }\textbf {\bibinfo {volume} {515}},\ \bibinfo
  {pages} {241} (\bibinfo {year} {2014})}\BibitemShut {NoStop}%
\bibitem [{\citenamefont {Leykam}\ \emph {et~al.}(2016)\citenamefont {Leykam},
  \citenamefont {Rechtsman},\ and\ \citenamefont
  {Chong}}]{PhysRevLett.117.013902}%
  \BibitemOpen
  \bibfield  {author} {\bibinfo {author} {\bibfnamefont {D.}~\bibnamefont
  {Leykam}}, \bibinfo {author} {\bibfnamefont {M.~C.}\ \bibnamefont
  {Rechtsman}},\ and\ \bibinfo {author} {\bibfnamefont {Y.~D.}\ \bibnamefont
  {Chong}},\ }\bibfield  {title} {\bibinfo {title} {Anomalous topological
  phases and unpaired dirac cones in photonic floquet topological insulators},\
  }\href {https://doi.org/10.1103/PhysRevLett.117.013902} {\bibfield  {journal}
  {\bibinfo  {journal} {Phys. Rev. Lett.}\ }\textbf {\bibinfo {volume} {117}},\
  \bibinfo {pages} {013902} (\bibinfo {year} {2016})}\BibitemShut {NoStop}%
\bibitem [{\citenamefont {R{\"u}ter}\ \emph {et~al.}(2010)\citenamefont
  {R{\"u}ter}, \citenamefont {Makris}, \citenamefont {El-Ganainy},
  \citenamefont {Christodoulides}, \citenamefont {Segev},\ and\ \citenamefont
  {Kip}}]{ruter2010observation}%
  \BibitemOpen
  \bibfield  {author} {\bibinfo {author} {\bibfnamefont {C.~E.}\ \bibnamefont
  {R{\"u}ter}}, \bibinfo {author} {\bibfnamefont {K.~G.}\ \bibnamefont
  {Makris}}, \bibinfo {author} {\bibfnamefont {R.}~\bibnamefont {El-Ganainy}},
  \bibinfo {author} {\bibfnamefont {D.~N.}\ \bibnamefont {Christodoulides}},
  \bibinfo {author} {\bibfnamefont {M.}~\bibnamefont {Segev}},\ and\ \bibinfo
  {author} {\bibfnamefont {D.}~\bibnamefont {Kip}},\ }\bibfield  {title}
  {\bibinfo {title} {Observation of parity--time symmetry in optics},\
  }\href@noop {} {\bibfield  {journal} {\bibinfo  {journal} {Nature physics}\
  }\textbf {\bibinfo {volume} {6}},\ \bibinfo {pages} {192} (\bibinfo {year}
  {2010})}\BibitemShut {NoStop}%
\bibitem [{\citenamefont {Feng}\ \emph {et~al.}(2017)\citenamefont {Feng},
  \citenamefont {El-Ganainy},\ and\ \citenamefont {Ge}}]{feng2017non}%
  \BibitemOpen
  \bibfield  {author} {\bibinfo {author} {\bibfnamefont {L.}~\bibnamefont
  {Feng}}, \bibinfo {author} {\bibfnamefont {R.}~\bibnamefont {El-Ganainy}},\
  and\ \bibinfo {author} {\bibfnamefont {L.}~\bibnamefont {Ge}},\ }\bibfield
  {title} {\bibinfo {title} {Non-hermitian photonics based on parity--time
  symmetry},\ }\href@noop {} {\bibfield  {journal} {\bibinfo  {journal} {Nature
  Photonics}\ }\textbf {\bibinfo {volume} {11}},\ \bibinfo {pages} {752}
  (\bibinfo {year} {2017})}\BibitemShut {NoStop}%
\bibitem [{\citenamefont {Zhou}\ and\ \citenamefont
  {Yu}(2019)}]{PhysRevA.99.043412}%
  \BibitemOpen
  \bibfield  {author} {\bibinfo {author} {\bibfnamefont {Z.}~\bibnamefont
  {Zhou}}\ and\ \bibinfo {author} {\bibfnamefont {Z.}~\bibnamefont {Yu}},\
  }\bibfield  {title} {\bibinfo {title} {Interaction effects on the
  $\mathcal{PT}$-symmetry-breaking transition in atomic gases},\ }\href
  {https://doi.org/10.1103/PhysRevA.99.043412} {\bibfield  {journal} {\bibinfo
  {journal} {Phys. Rev. A}\ }\textbf {\bibinfo {volume} {99}},\ \bibinfo
  {pages} {043412} (\bibinfo {year} {2019})}\BibitemShut {NoStop}%
\bibitem [{\citenamefont {Bender}\ and\ \citenamefont
  {Boettcher}(1998)}]{PhysRevLett.80.5243}%
  \BibitemOpen
  \bibfield  {author} {\bibinfo {author} {\bibfnamefont {C.~M.}\ \bibnamefont
  {Bender}}\ and\ \bibinfo {author} {\bibfnamefont {S.}~\bibnamefont
  {Boettcher}},\ }\bibfield  {title} {\bibinfo {title} {Real spectra in
  non-hermitian hamiltonians having $\mathcal{P}\mathcal{T}$ symmetry},\ }\href
  {https://doi.org/10.1103/PhysRevLett.80.5243} {\bibfield  {journal} {\bibinfo
   {journal} {Phys. Rev. Lett.}\ }\textbf {\bibinfo {volume} {80}},\ \bibinfo
  {pages} {5243} (\bibinfo {year} {1998})}\BibitemShut {NoStop}%
\bibitem [{\citenamefont {Liu}\ \emph {et~al.}(2024{\natexlab{a}})\citenamefont
  {Liu}, \citenamefont {Liu}, \citenamefont {Ni}, \citenamefont {Jia},
  \citenamefont {Ziegler}, \citenamefont {Al{\`u}},\ and\ \citenamefont
  {Chen}}]{liu2024floquet}%
  \BibitemOpen
  \bibfield  {author} {\bibinfo {author} {\bibfnamefont {W.}~\bibnamefont
  {Liu}}, \bibinfo {author} {\bibfnamefont {Q.}~\bibnamefont {Liu}}, \bibinfo
  {author} {\bibfnamefont {X.}~\bibnamefont {Ni}}, \bibinfo {author}
  {\bibfnamefont {Y.}~\bibnamefont {Jia}}, \bibinfo {author} {\bibfnamefont
  {K.}~\bibnamefont {Ziegler}}, \bibinfo {author} {\bibfnamefont
  {A.}~\bibnamefont {Al{\`u}}},\ and\ \bibinfo {author} {\bibfnamefont
  {F.}~\bibnamefont {Chen}},\ }\bibfield  {title} {\bibinfo {title} {Floquet
  parity-time symmetry in integrated photonics},\ }\href@noop {} {\bibfield
  {journal} {\bibinfo  {journal} {Nature Communications}\ }\textbf {\bibinfo
  {volume} {15}},\ \bibinfo {pages} {946} (\bibinfo {year}
  {2024}{\natexlab{a}})}\BibitemShut {NoStop}%
\bibitem [{\citenamefont {Xia}\ \emph {et~al.}(2021)\citenamefont {Xia},
  \citenamefont {Kaltsas}, \citenamefont {Song}, \citenamefont {Komis},
  \citenamefont {Xu}, \citenamefont {Szameit}, \citenamefont {Buljan},
  \citenamefont {Makris},\ and\ \citenamefont {Chen}}]{science.abf6873}%
  \BibitemOpen
  \bibfield  {author} {\bibinfo {author} {\bibfnamefont {S.}~\bibnamefont
  {Xia}}, \bibinfo {author} {\bibfnamefont {D.}~\bibnamefont {Kaltsas}},
  \bibinfo {author} {\bibfnamefont {D.}~\bibnamefont {Song}}, \bibinfo {author}
  {\bibfnamefont {I.}~\bibnamefont {Komis}}, \bibinfo {author} {\bibfnamefont
  {J.}~\bibnamefont {Xu}}, \bibinfo {author} {\bibfnamefont {A.}~\bibnamefont
  {Szameit}}, \bibinfo {author} {\bibfnamefont {H.}~\bibnamefont {Buljan}},
  \bibinfo {author} {\bibfnamefont {K.~G.}\ \bibnamefont {Makris}},\ and\
  \bibinfo {author} {\bibfnamefont {Z.}~\bibnamefont {Chen}},\ }\bibfield
  {title} {\bibinfo {title} {Nonlinear tuning of pt symmetry and non-hermitian
  topological states},\ }\href {https://doi.org/10.1126/science.abf6873}
  {\bibfield  {journal} {\bibinfo  {journal} {Science}\ }\textbf {\bibinfo
  {volume} {372}},\ \bibinfo {pages} {72} (\bibinfo {year} {2021})},\ \Eprint
  {https://arxiv.org/abs/https://www.science.org/doi/pdf/10.1126/science.abf6873}
  {https://www.science.org/doi/pdf/10.1126/science.abf6873} \BibitemShut
  {NoStop}%
\bibitem [{\citenamefont {Shi}\ \emph {et~al.}(2016)\citenamefont {Shi},
  \citenamefont {Dubois}, \citenamefont {Chen}, \citenamefont {Cheng},
  \citenamefont {Ramezani}, \citenamefont {Wang},\ and\ \citenamefont
  {Zhang}}]{shi2016accessing}%
  \BibitemOpen
  \bibfield  {author} {\bibinfo {author} {\bibfnamefont {C.}~\bibnamefont
  {Shi}}, \bibinfo {author} {\bibfnamefont {M.}~\bibnamefont {Dubois}},
  \bibinfo {author} {\bibfnamefont {Y.}~\bibnamefont {Chen}}, \bibinfo {author}
  {\bibfnamefont {L.}~\bibnamefont {Cheng}}, \bibinfo {author} {\bibfnamefont
  {H.}~\bibnamefont {Ramezani}}, \bibinfo {author} {\bibfnamefont
  {Y.}~\bibnamefont {Wang}},\ and\ \bibinfo {author} {\bibfnamefont
  {X.}~\bibnamefont {Zhang}},\ }\bibfield  {title} {\bibinfo {title} {Accessing
  the exceptional points of parity-time symmetric acoustics},\ }\href@noop {}
  {\bibfield  {journal} {\bibinfo  {journal} {Nature communications}\ }\textbf
  {\bibinfo {volume} {7}},\ \bibinfo {pages} {11110} (\bibinfo {year}
  {2016})}\BibitemShut {NoStop}%
\bibitem [{\citenamefont {Cao}\ \emph {et~al.}(2022)\citenamefont {Cao},
  \citenamefont {Wang}, \citenamefont {Chen}, \citenamefont {Hu}, \citenamefont
  {Wang}, \citenamefont {Yang},\ and\ \citenamefont {Zhang}}]{cao2022fully}%
  \BibitemOpen
  \bibfield  {author} {\bibinfo {author} {\bibfnamefont {W.}~\bibnamefont
  {Cao}}, \bibinfo {author} {\bibfnamefont {C.}~\bibnamefont {Wang}}, \bibinfo
  {author} {\bibfnamefont {W.}~\bibnamefont {Chen}}, \bibinfo {author}
  {\bibfnamefont {S.}~\bibnamefont {Hu}}, \bibinfo {author} {\bibfnamefont
  {H.}~\bibnamefont {Wang}}, \bibinfo {author} {\bibfnamefont {L.}~\bibnamefont
  {Yang}},\ and\ \bibinfo {author} {\bibfnamefont {X.}~\bibnamefont {Zhang}},\
  }\bibfield  {title} {\bibinfo {title} {Fully integrated
  parity--time-symmetric electronics},\ }\href@noop {} {\bibfield  {journal}
  {\bibinfo  {journal} {Nature nanotechnology}\ }\textbf {\bibinfo {volume}
  {17}},\ \bibinfo {pages} {262} (\bibinfo {year} {2022})}\BibitemShut
  {NoStop}%
\bibitem [{\citenamefont {Liu}\ \emph {et~al.}(2024{\natexlab{b}})\citenamefont
  {Liu}, \citenamefont {Mandal}, \citenamefont {Zhou}, \citenamefont {Xi},
  \citenamefont {Banerjee}, \citenamefont {Hu}, \citenamefont {Wei},
  \citenamefont {Wang}, \citenamefont {Wang}, \citenamefont {Gao},
  \citenamefont {Chen}, \citenamefont {Yang}, \citenamefont {Chong},\ and\
  \citenamefont {Zhang}}]{PhysRevLett.132.113802}%
  \BibitemOpen
  \bibfield  {author} {\bibinfo {author} {\bibfnamefont {G.-G.}\ \bibnamefont
  {Liu}}, \bibinfo {author} {\bibfnamefont {S.}~\bibnamefont {Mandal}},
  \bibinfo {author} {\bibfnamefont {P.}~\bibnamefont {Zhou}}, \bibinfo {author}
  {\bibfnamefont {X.}~\bibnamefont {Xi}}, \bibinfo {author} {\bibfnamefont
  {R.}~\bibnamefont {Banerjee}}, \bibinfo {author} {\bibfnamefont {Y.-H.}\
  \bibnamefont {Hu}}, \bibinfo {author} {\bibfnamefont {M.}~\bibnamefont
  {Wei}}, \bibinfo {author} {\bibfnamefont {M.}~\bibnamefont {Wang}}, \bibinfo
  {author} {\bibfnamefont {Q.}~\bibnamefont {Wang}}, \bibinfo {author}
  {\bibfnamefont {Z.}~\bibnamefont {Gao}}, \bibinfo {author} {\bibfnamefont
  {H.}~\bibnamefont {Chen}}, \bibinfo {author} {\bibfnamefont {Y.}~\bibnamefont
  {Yang}}, \bibinfo {author} {\bibfnamefont {Y.}~\bibnamefont {Chong}},\ and\
  \bibinfo {author} {\bibfnamefont {B.}~\bibnamefont {Zhang}},\ }\bibfield
  {title} {\bibinfo {title} {Localization of chiral edge states by the
  non-hermitian skin effect},\ }\href
  {https://doi.org/10.1103/PhysRevLett.132.113802} {\bibfield  {journal}
  {\bibinfo  {journal} {Phys. Rev. Lett.}\ }\textbf {\bibinfo {volume} {132}},\
  \bibinfo {pages} {113802} (\bibinfo {year} {2024}{\natexlab{b}})}\BibitemShut
  {NoStop}%
\bibitem [{\citenamefont {{\c{S}}eker}\ \emph {et~al.}(2023)\citenamefont
  {{\c{S}}eker}, \citenamefont {Olyaeefar}, \citenamefont {Dadashi},
  \citenamefont {{\c{S}}eng{\"u}l}, \citenamefont {Teimourpour}, \citenamefont
  {El-Ganainy},\ and\ \citenamefont {Demir}}]{cseker2023single}%
  \BibitemOpen
  \bibfield  {author} {\bibinfo {author} {\bibfnamefont {E.}~\bibnamefont
  {{\c{S}}eker}}, \bibinfo {author} {\bibfnamefont {B.}~\bibnamefont
  {Olyaeefar}}, \bibinfo {author} {\bibfnamefont {K.}~\bibnamefont {Dadashi}},
  \bibinfo {author} {\bibfnamefont {S.}~\bibnamefont {{\c{S}}eng{\"u}l}},
  \bibinfo {author} {\bibfnamefont {M.~H.}\ \bibnamefont {Teimourpour}},
  \bibinfo {author} {\bibfnamefont {R.}~\bibnamefont {El-Ganainy}},\ and\
  \bibinfo {author} {\bibfnamefont {A.}~\bibnamefont {Demir}},\ }\bibfield
  {title} {\bibinfo {title} {Single-mode quasi pt-symmetric laser with high
  power emission},\ }\href@noop {} {\bibfield  {journal} {\bibinfo  {journal}
  {Light: Science \& Applications}\ }\textbf {\bibinfo {volume} {12}},\
  \bibinfo {pages} {149} (\bibinfo {year} {2023})}\BibitemShut {NoStop}%
\bibitem [{\citenamefont {Ezawa}(2021)}]{PhysRevResearch.3.043006}%
  \BibitemOpen
  \bibfield  {author} {\bibinfo {author} {\bibfnamefont {M.}~\bibnamefont
  {Ezawa}},\ }\bibfield  {title} {\bibinfo {title} {Non-hermitian non-abelian
  topological insulators with $\mathcal{PT}$ symmetry},\ }\href
  {https://doi.org/10.1103/PhysRevResearch.3.043006} {\bibfield  {journal}
  {\bibinfo  {journal} {Phys. Rev. Res.}\ }\textbf {\bibinfo {volume} {3}},\
  \bibinfo {pages} {043006} (\bibinfo {year} {2021})}\BibitemShut {NoStop}%
\bibitem [{\citenamefont {Wimmer}\ \emph {et~al.}(2015)\citenamefont {Wimmer},
  \citenamefont {Regensburger}, \citenamefont {Miri}, \citenamefont {Bersch},
  \citenamefont {Christodoulides},\ and\ \citenamefont
  {Peschel}}]{wimmer2015observation}%
  \BibitemOpen
  \bibfield  {author} {\bibinfo {author} {\bibfnamefont {M.}~\bibnamefont
  {Wimmer}}, \bibinfo {author} {\bibfnamefont {A.}~\bibnamefont
  {Regensburger}}, \bibinfo {author} {\bibfnamefont {M.-A.}\ \bibnamefont
  {Miri}}, \bibinfo {author} {\bibfnamefont {C.}~\bibnamefont {Bersch}},
  \bibinfo {author} {\bibfnamefont {D.~N.}\ \bibnamefont {Christodoulides}},\
  and\ \bibinfo {author} {\bibfnamefont {U.}~\bibnamefont {Peschel}},\
  }\bibfield  {title} {\bibinfo {title} {Observation of optical solitons in
  pt-symmetric lattices},\ }\href@noop {} {\bibfield  {journal} {\bibinfo
  {journal} {Nature communications}\ }\textbf {\bibinfo {volume} {6}},\
  \bibinfo {pages} {7782} (\bibinfo {year} {2015})}\BibitemShut {NoStop}%
\bibitem [{\citenamefont {Xu}\ \emph {et~al.}(2016)\citenamefont {Xu},
  \citenamefont {Fegadolli}, \citenamefont {Gan}, \citenamefont {Lu},
  \citenamefont {Liu}, \citenamefont {Li}, \citenamefont {Scherer},\ and\
  \citenamefont {Chen}}]{xu2016experimental}%
  \BibitemOpen
  \bibfield  {author} {\bibinfo {author} {\bibfnamefont {Y.-L.}\ \bibnamefont
  {Xu}}, \bibinfo {author} {\bibfnamefont {W.~S.}\ \bibnamefont {Fegadolli}},
  \bibinfo {author} {\bibfnamefont {L.}~\bibnamefont {Gan}}, \bibinfo {author}
  {\bibfnamefont {M.-H.}\ \bibnamefont {Lu}}, \bibinfo {author} {\bibfnamefont
  {X.-P.}\ \bibnamefont {Liu}}, \bibinfo {author} {\bibfnamefont {Z.-Y.}\
  \bibnamefont {Li}}, \bibinfo {author} {\bibfnamefont {A.}~\bibnamefont
  {Scherer}},\ and\ \bibinfo {author} {\bibfnamefont {Y.-F.}\ \bibnamefont
  {Chen}},\ }\bibfield  {title} {\bibinfo {title} {Experimental realization of
  bloch oscillations in a parity-time synthetic silicon photonic lattice},\
  }\href@noop {} {\bibfield  {journal} {\bibinfo  {journal} {Nature
  communications}\ }\textbf {\bibinfo {volume} {7}},\ \bibinfo {pages} {11319}
  (\bibinfo {year} {2016})}\BibitemShut {NoStop}%
\bibitem [{\citenamefont {Yin}\ \emph {et~al.}(2022)\citenamefont {Yin},
  \citenamefont {Galiffi},\ and\ \citenamefont {Al{\`u}}}]{yin2022floquet}%
  \BibitemOpen
  \bibfield  {author} {\bibinfo {author} {\bibfnamefont {S.}~\bibnamefont
  {Yin}}, \bibinfo {author} {\bibfnamefont {E.}~\bibnamefont {Galiffi}},\ and\
  \bibinfo {author} {\bibfnamefont {A.}~\bibnamefont {Al{\`u}}},\ }\bibfield
  {title} {\bibinfo {title} {Floquet metamaterials},\ }\href@noop {} {\bibfield
   {journal} {\bibinfo  {journal} {ELight}\ }\textbf {\bibinfo {volume} {2}},\
  \bibinfo {pages} {8} (\bibinfo {year} {2022})}\BibitemShut {NoStop}%
\bibitem [{\citenamefont {Lin}\ \emph {et~al.}(2016)\citenamefont {Lin},
  \citenamefont {Xiao}, \citenamefont {Yuan},\ and\ \citenamefont
  {Fan}}]{lin2016photonic}%
  \BibitemOpen
  \bibfield  {author} {\bibinfo {author} {\bibfnamefont {Q.}~\bibnamefont
  {Lin}}, \bibinfo {author} {\bibfnamefont {M.}~\bibnamefont {Xiao}}, \bibinfo
  {author} {\bibfnamefont {L.}~\bibnamefont {Yuan}},\ and\ \bibinfo {author}
  {\bibfnamefont {S.}~\bibnamefont {Fan}},\ }\bibfield  {title} {\bibinfo
  {title} {Photonic weyl point in a two-dimensional resonator lattice with a
  synthetic frequency dimension},\ }\href@noop {} {\bibfield  {journal}
  {\bibinfo  {journal} {Nature communications}\ }\textbf {\bibinfo {volume}
  {7}},\ \bibinfo {pages} {13731} (\bibinfo {year} {2016})}\BibitemShut
  {NoStop}%
\bibitem [{\citenamefont {Li}\ \emph {et~al.}(2023{\natexlab{a}})\citenamefont
  {Li}, \citenamefont {Wang}, \citenamefont {Ye}, \citenamefont {Zheng},
  \citenamefont {Wang}, \citenamefont {Liu}, \citenamefont {Dutt},
  \citenamefont {Yuan},\ and\ \citenamefont {Chen}}]{li2023direct}%
  \BibitemOpen
  \bibfield  {author} {\bibinfo {author} {\bibfnamefont {G.}~\bibnamefont
  {Li}}, \bibinfo {author} {\bibfnamefont {L.}~\bibnamefont {Wang}}, \bibinfo
  {author} {\bibfnamefont {R.}~\bibnamefont {Ye}}, \bibinfo {author}
  {\bibfnamefont {Y.}~\bibnamefont {Zheng}}, \bibinfo {author} {\bibfnamefont
  {D.-W.}\ \bibnamefont {Wang}}, \bibinfo {author} {\bibfnamefont {X.-J.}\
  \bibnamefont {Liu}}, \bibinfo {author} {\bibfnamefont {A.}~\bibnamefont
  {Dutt}}, \bibinfo {author} {\bibfnamefont {L.}~\bibnamefont {Yuan}},\ and\
  \bibinfo {author} {\bibfnamefont {X.}~\bibnamefont {Chen}},\ }\bibfield
  {title} {\bibinfo {title} {Direct extraction of topological zak phase with
  the synthetic dimension},\ }\href@noop {} {\bibfield  {journal} {\bibinfo
  {journal} {Light: Science \& Applications}\ }\textbf {\bibinfo {volume}
  {12}},\ \bibinfo {pages} {81} (\bibinfo {year}
  {2023}{\natexlab{a}})}\BibitemShut {NoStop}%
\bibitem [{\citenamefont {Cheng}\ \emph {et~al.}(2023)\citenamefont {Cheng},
  \citenamefont {Lustig}, \citenamefont {Wang},\ and\ \citenamefont
  {Fan}}]{cheng2023multi}%
  \BibitemOpen
  \bibfield  {author} {\bibinfo {author} {\bibfnamefont {D.}~\bibnamefont
  {Cheng}}, \bibinfo {author} {\bibfnamefont {E.}~\bibnamefont {Lustig}},
  \bibinfo {author} {\bibfnamefont {K.}~\bibnamefont {Wang}},\ and\ \bibinfo
  {author} {\bibfnamefont {S.}~\bibnamefont {Fan}},\ }\bibfield  {title}
  {\bibinfo {title} {Multi-dimensional band structure spectroscopy in the
  synthetic frequency dimension},\ }\href@noop {} {\bibfield  {journal}
  {\bibinfo  {journal} {Light: Science \& Applications}\ }\textbf {\bibinfo
  {volume} {12}},\ \bibinfo {pages} {158} (\bibinfo {year} {2023})}\BibitemShut
  {NoStop}%
\bibitem [{\citenamefont {Vezzoli}\ \emph {et~al.}(2018)\citenamefont
  {Vezzoli}, \citenamefont {Bruno}, \citenamefont {DeVault}, \citenamefont
  {Roger}, \citenamefont {Shalaev}, \citenamefont {Boltasseva}, \citenamefont
  {Ferrera}, \citenamefont {Clerici}, \citenamefont {Dubietis},\ and\
  \citenamefont {Faccio}}]{PhysRevLett.120.043902}%
  \BibitemOpen
  \bibfield  {author} {\bibinfo {author} {\bibfnamefont {S.}~\bibnamefont
  {Vezzoli}}, \bibinfo {author} {\bibfnamefont {V.}~\bibnamefont {Bruno}},
  \bibinfo {author} {\bibfnamefont {C.}~\bibnamefont {DeVault}}, \bibinfo
  {author} {\bibfnamefont {T.}~\bibnamefont {Roger}}, \bibinfo {author}
  {\bibfnamefont {V.~M.}\ \bibnamefont {Shalaev}}, \bibinfo {author}
  {\bibfnamefont {A.}~\bibnamefont {Boltasseva}}, \bibinfo {author}
  {\bibfnamefont {M.}~\bibnamefont {Ferrera}}, \bibinfo {author} {\bibfnamefont
  {M.}~\bibnamefont {Clerici}}, \bibinfo {author} {\bibfnamefont
  {A.}~\bibnamefont {Dubietis}},\ and\ \bibinfo {author} {\bibfnamefont
  {D.}~\bibnamefont {Faccio}},\ }\bibfield  {title} {\bibinfo {title} {Optical
  time reversal from time-dependent epsilon-near-zero media},\ }\href
  {https://doi.org/10.1103/PhysRevLett.120.043902} {\bibfield  {journal}
  {\bibinfo  {journal} {Phys. Rev. Lett.}\ }\textbf {\bibinfo {volume} {120}},\
  \bibinfo {pages} {043902} (\bibinfo {year} {2018})}\BibitemShut {NoStop}%
\bibitem [{\citenamefont {Peng}(2022)}]{PhysRevLett.128.186802}%
  \BibitemOpen
  \bibfield  {author} {\bibinfo {author} {\bibfnamefont {Y.}~\bibnamefont
  {Peng}},\ }\bibfield  {title} {\bibinfo {title} {Topological space-time
  crystal},\ }\href {https://doi.org/10.1103/PhysRevLett.128.186802} {\bibfield
   {journal} {\bibinfo  {journal} {Phys. Rev. Lett.}\ }\textbf {\bibinfo
  {volume} {128}},\ \bibinfo {pages} {186802} (\bibinfo {year}
  {2022})}\BibitemShut {NoStop}%
\bibitem [{\citenamefont {Dong}\ \emph {et~al.}(2024)\citenamefont {Dong},
  \citenamefont {Li}, \citenamefont {Wan}, \citenamefont {Liang}, \citenamefont
  {Yang},\ and\ \citenamefont {Yan}}]{dong2024quantum}%
  \BibitemOpen
  \bibfield  {author} {\bibinfo {author} {\bibfnamefont {Z.}~\bibnamefont
  {Dong}}, \bibinfo {author} {\bibfnamefont {H.}~\bibnamefont {Li}}, \bibinfo
  {author} {\bibfnamefont {T.}~\bibnamefont {Wan}}, \bibinfo {author}
  {\bibfnamefont {Q.}~\bibnamefont {Liang}}, \bibinfo {author} {\bibfnamefont
  {Z.}~\bibnamefont {Yang}},\ and\ \bibinfo {author} {\bibfnamefont
  {B.}~\bibnamefont {Yan}},\ }\bibfield  {title} {\bibinfo {title} {Quantum
  time reflection and refraction of ultracold atoms},\ }\href@noop {}
  {\bibfield  {journal} {\bibinfo  {journal} {Nature Photonics}\ }\textbf
  {\bibinfo {volume} {18}},\ \bibinfo {pages} {68} (\bibinfo {year}
  {2024})}\BibitemShut {NoStop}%
\bibitem [{\citenamefont {Else}\ \emph {et~al.}(2016)\citenamefont {Else},
  \citenamefont {Bauer},\ and\ \citenamefont {Nayak}}]{PhysRevLett.117.090402}%
  \BibitemOpen
  \bibfield  {author} {\bibinfo {author} {\bibfnamefont {D.~V.}\ \bibnamefont
  {Else}}, \bibinfo {author} {\bibfnamefont {B.}~\bibnamefont {Bauer}},\ and\
  \bibinfo {author} {\bibfnamefont {C.}~\bibnamefont {Nayak}},\ }\bibfield
  {title} {\bibinfo {title} {Floquet time crystals},\ }\href
  {https://doi.org/10.1103/PhysRevLett.117.090402} {\bibfield  {journal}
  {\bibinfo  {journal} {Phys. Rev. Lett.}\ }\textbf {\bibinfo {volume} {117}},\
  \bibinfo {pages} {090402} (\bibinfo {year} {2016})}\BibitemShut {NoStop}%
\bibitem [{\citenamefont {Eckardt}(2017)}]{RevModPhys.89.011004}%
  \BibitemOpen
  \bibfield  {author} {\bibinfo {author} {\bibfnamefont {A.}~\bibnamefont
  {Eckardt}},\ }\bibfield  {title} {\bibinfo {title} {Colloquium: Atomic
  quantum gases in periodically driven optical lattices},\ }\href
  {https://doi.org/10.1103/RevModPhys.89.011004} {\bibfield  {journal}
  {\bibinfo  {journal} {Rev. Mod. Phys.}\ }\textbf {\bibinfo {volume} {89}},\
  \bibinfo {pages} {011004} (\bibinfo {year} {2017})}\BibitemShut {NoStop}%
\bibitem [{\citenamefont {Geier}\ \emph {et~al.}(2021)\citenamefont {Geier},
  \citenamefont {Thaicharoen}, \citenamefont {Hainaut}, \citenamefont {Franz},
  \citenamefont {Salzinger}, \citenamefont {Tebben}, \citenamefont
  {Grimshandl}, \citenamefont {Z{\"u}rn},\ and\ \citenamefont
  {Weidem{\"u}ller}}]{geier2021floquet}%
  \BibitemOpen
  \bibfield  {author} {\bibinfo {author} {\bibfnamefont {S.}~\bibnamefont
  {Geier}}, \bibinfo {author} {\bibfnamefont {N.}~\bibnamefont {Thaicharoen}},
  \bibinfo {author} {\bibfnamefont {C.}~\bibnamefont {Hainaut}}, \bibinfo
  {author} {\bibfnamefont {T.}~\bibnamefont {Franz}}, \bibinfo {author}
  {\bibfnamefont {A.}~\bibnamefont {Salzinger}}, \bibinfo {author}
  {\bibfnamefont {A.}~\bibnamefont {Tebben}}, \bibinfo {author} {\bibfnamefont
  {D.}~\bibnamefont {Grimshandl}}, \bibinfo {author} {\bibfnamefont
  {G.}~\bibnamefont {Z{\"u}rn}},\ and\ \bibinfo {author} {\bibfnamefont
  {M.}~\bibnamefont {Weidem{\"u}ller}},\ }\bibfield  {title} {\bibinfo {title}
  {Floquet hamiltonian engineering of an isolated many-body spin system},\
  }\href@noop {} {\bibfield  {journal} {\bibinfo  {journal} {Science}\ }\textbf
  {\bibinfo {volume} {374}},\ \bibinfo {pages} {1149} (\bibinfo {year}
  {2021})}\BibitemShut {NoStop}%
\bibitem [{\citenamefont {Li}\ \emph {et~al.}(2023{\natexlab{b}})\citenamefont
  {Li}, \citenamefont {Li}, \citenamefont {Yan}, \citenamefont {Li},
  \citenamefont {Gong},\ and\ \citenamefont {Li}}]{li2023fractal}%
  \BibitemOpen
  \bibfield  {author} {\bibinfo {author} {\bibfnamefont {M.}~\bibnamefont
  {Li}}, \bibinfo {author} {\bibfnamefont {C.}~\bibnamefont {Li}}, \bibinfo
  {author} {\bibfnamefont {L.}~\bibnamefont {Yan}}, \bibinfo {author}
  {\bibfnamefont {Q.}~\bibnamefont {Li}}, \bibinfo {author} {\bibfnamefont
  {Q.}~\bibnamefont {Gong}},\ and\ \bibinfo {author} {\bibfnamefont
  {Y.}~\bibnamefont {Li}},\ }\bibfield  {title} {\bibinfo {title} {Fractal
  photonic anomalous floquet topological insulators to generate multiple
  quantum chiral edge states},\ }\href@noop {} {\bibfield  {journal} {\bibinfo
  {journal} {Light: Science \& Applications}\ }\textbf {\bibinfo {volume}
  {12}},\ \bibinfo {pages} {262} (\bibinfo {year}
  {2023}{\natexlab{b}})}\BibitemShut {NoStop}%
\bibitem [{\citenamefont {Xu}\ and\ \citenamefont
  {Wu}(2018)}]{PhysRevLett.120.096401}%
  \BibitemOpen
  \bibfield  {author} {\bibinfo {author} {\bibfnamefont {S.}~\bibnamefont
  {Xu}}\ and\ \bibinfo {author} {\bibfnamefont {C.}~\bibnamefont {Wu}},\
  }\bibfield  {title} {\bibinfo {title} {Space-time crystal and space-time
  group},\ }\href {https://doi.org/10.1103/PhysRevLett.120.096401} {\bibfield
  {journal} {\bibinfo  {journal} {Phys. Rev. Lett.}\ }\textbf {\bibinfo
  {volume} {120}},\ \bibinfo {pages} {096401} (\bibinfo {year}
  {2018})}\BibitemShut {NoStop}%
\bibitem [{\citenamefont {Pan}\ \emph {et~al.}(2023)\citenamefont {Pan},
  \citenamefont {Cohen},\ and\ \citenamefont {Segev}}]{PhysRevLett.130.233801}%
  \BibitemOpen
  \bibfield  {author} {\bibinfo {author} {\bibfnamefont {Y.}~\bibnamefont
  {Pan}}, \bibinfo {author} {\bibfnamefont {M.-I.}\ \bibnamefont {Cohen}},\
  and\ \bibinfo {author} {\bibfnamefont {M.}~\bibnamefont {Segev}},\ }\bibfield
   {title} {\bibinfo {title} {Superluminal $k$-gap solitons in nonlinear
  photonic time crystals},\ }\href
  {https://doi.org/10.1103/PhysRevLett.130.233801} {\bibfield  {journal}
  {\bibinfo  {journal} {Phys. Rev. Lett.}\ }\textbf {\bibinfo {volume} {130}},\
  \bibinfo {pages} {233801} (\bibinfo {year} {2023})}\BibitemShut {NoStop}%
\bibitem [{\citenamefont {Zhou}\ \emph {et~al.}(2023)\citenamefont {Zhou},
  \citenamefont {Bao}, \citenamefont {Fan}, \citenamefont {Zhou}, \citenamefont
  {Gao}, \citenamefont {Zhong}, \citenamefont {Lin}, \citenamefont {Liu},
  \citenamefont {Yu}, \citenamefont {Tang} \emph
  {et~al.}}]{zhou2023pseudospin}%
  \BibitemOpen
  \bibfield  {author} {\bibinfo {author} {\bibfnamefont {S.}~\bibnamefont
  {Zhou}}, \bibinfo {author} {\bibfnamefont {C.}~\bibnamefont {Bao}}, \bibinfo
  {author} {\bibfnamefont {B.}~\bibnamefont {Fan}}, \bibinfo {author}
  {\bibfnamefont {H.}~\bibnamefont {Zhou}}, \bibinfo {author} {\bibfnamefont
  {Q.}~\bibnamefont {Gao}}, \bibinfo {author} {\bibfnamefont {H.}~\bibnamefont
  {Zhong}}, \bibinfo {author} {\bibfnamefont {T.}~\bibnamefont {Lin}}, \bibinfo
  {author} {\bibfnamefont {H.}~\bibnamefont {Liu}}, \bibinfo {author}
  {\bibfnamefont {P.}~\bibnamefont {Yu}}, \bibinfo {author} {\bibfnamefont
  {P.}~\bibnamefont {Tang}}, \emph {et~al.},\ }\bibfield  {title} {\bibinfo
  {title} {Pseudospin-selective floquet band engineering in black phosphorus},\
  }\href@noop {} {\bibfield  {journal} {\bibinfo  {journal} {Nature}\ }\textbf
  {\bibinfo {volume} {614}},\ \bibinfo {pages} {75} (\bibinfo {year}
  {2023})}\BibitemShut {NoStop}%
\bibitem [{\citenamefont {Lindner}\ \emph {et~al.}(2011)\citenamefont
  {Lindner}, \citenamefont {Refael},\ and\ \citenamefont
  {Galitski}}]{lindner2011floquet}%
  \BibitemOpen
  \bibfield  {author} {\bibinfo {author} {\bibfnamefont {N.~H.}\ \bibnamefont
  {Lindner}}, \bibinfo {author} {\bibfnamefont {G.}~\bibnamefont {Refael}},\
  and\ \bibinfo {author} {\bibfnamefont {V.}~\bibnamefont {Galitski}},\
  }\bibfield  {title} {\bibinfo {title} {Floquet topological insulator in
  semiconductor quantum wells},\ }\href@noop {} {\bibfield  {journal} {\bibinfo
   {journal} {Nature Physics}\ }\textbf {\bibinfo {volume} {7}},\ \bibinfo
  {pages} {490} (\bibinfo {year} {2011})}\BibitemShut {NoStop}%
\bibitem [{\citenamefont {Lustig}\ \emph {et~al.}(2018)\citenamefont {Lustig},
  \citenamefont {Sharabi},\ and\ \citenamefont
  {Segev}}]{lustig2018topological}%
  \BibitemOpen
  \bibfield  {author} {\bibinfo {author} {\bibfnamefont {E.}~\bibnamefont
  {Lustig}}, \bibinfo {author} {\bibfnamefont {Y.}~\bibnamefont {Sharabi}},\
  and\ \bibinfo {author} {\bibfnamefont {M.}~\bibnamefont {Segev}},\ }\bibfield
   {title} {\bibinfo {title} {Topological aspects of photonic time crystals},\
  }\href@noop {} {\bibfield  {journal} {\bibinfo  {journal} {Optica}\ }\textbf
  {\bibinfo {volume} {5}},\ \bibinfo {pages} {1390} (\bibinfo {year}
  {2018})}\BibitemShut {NoStop}%
\bibitem [{\citenamefont {Wang}\ \emph {et~al.}(2023)\citenamefont {Wang},
  \citenamefont {Mirmoosa}, \citenamefont {Asadchy}, \citenamefont {Rockstuhl},
  \citenamefont {Fan},\ and\ \citenamefont {Tretyakov}}]{wang2023metasurface}%
  \BibitemOpen
  \bibfield  {author} {\bibinfo {author} {\bibfnamefont {X.}~\bibnamefont
  {Wang}}, \bibinfo {author} {\bibfnamefont {M.~S.}\ \bibnamefont {Mirmoosa}},
  \bibinfo {author} {\bibfnamefont {V.~S.}\ \bibnamefont {Asadchy}}, \bibinfo
  {author} {\bibfnamefont {C.}~\bibnamefont {Rockstuhl}}, \bibinfo {author}
  {\bibfnamefont {S.}~\bibnamefont {Fan}},\ and\ \bibinfo {author}
  {\bibfnamefont {S.~A.}\ \bibnamefont {Tretyakov}},\ }\bibfield  {title}
  {\bibinfo {title} {Metasurface-based realization of photonic time crystals},\
  }\href@noop {} {\bibfield  {journal} {\bibinfo  {journal} {Science Advances}\
  }\textbf {\bibinfo {volume} {9}},\ \bibinfo {pages} {eadg7541} (\bibinfo
  {year} {2023})}\BibitemShut {NoStop}%
\bibitem [{\citenamefont {Lyubarov}\ \emph {et~al.}(2022)\citenamefont
  {Lyubarov}, \citenamefont {Lumer}, \citenamefont {Dikopoltsev}, \citenamefont
  {Lustig}, \citenamefont {Sharabi},\ and\ \citenamefont
  {Segev}}]{lyubarov2022amplified}%
  \BibitemOpen
  \bibfield  {author} {\bibinfo {author} {\bibfnamefont {M.}~\bibnamefont
  {Lyubarov}}, \bibinfo {author} {\bibfnamefont {Y.}~\bibnamefont {Lumer}},
  \bibinfo {author} {\bibfnamefont {A.}~\bibnamefont {Dikopoltsev}}, \bibinfo
  {author} {\bibfnamefont {E.}~\bibnamefont {Lustig}}, \bibinfo {author}
  {\bibfnamefont {Y.}~\bibnamefont {Sharabi}},\ and\ \bibinfo {author}
  {\bibfnamefont {M.}~\bibnamefont {Segev}},\ }\bibfield  {title} {\bibinfo
  {title} {Amplified emission and lasing in photonic time crystals},\
  }\href@noop {} {\bibfield  {journal} {\bibinfo  {journal} {Science}\ }\textbf
  {\bibinfo {volume} {377}},\ \bibinfo {pages} {425} (\bibinfo {year}
  {2022})}\BibitemShut {NoStop}%
\bibitem [{\citenamefont {Liu}\ \emph {et~al.}(2023)\citenamefont {Liu},
  \citenamefont {Ou}, \citenamefont {MacDonald},\ and\ \citenamefont
  {Zheludev}}]{liu2023photonic}%
  \BibitemOpen
  \bibfield  {author} {\bibinfo {author} {\bibfnamefont {T.}~\bibnamefont
  {Liu}}, \bibinfo {author} {\bibfnamefont {J.-Y.}\ \bibnamefont {Ou}},
  \bibinfo {author} {\bibfnamefont {K.~F.}\ \bibnamefont {MacDonald}},\ and\
  \bibinfo {author} {\bibfnamefont {N.~I.}\ \bibnamefont {Zheludev}},\
  }\bibfield  {title} {\bibinfo {title} {Photonic metamaterial analogue of a
  continuous time crystal},\ }\href@noop {} {\bibfield  {journal} {\bibinfo
  {journal} {Nature Physics}\ }\textbf {\bibinfo {volume} {19}},\ \bibinfo
  {pages} {986} (\bibinfo {year} {2023})}\BibitemShut {NoStop}%
\bibitem [{\citenamefont {Li}\ \emph {et~al.}(2020)\citenamefont {Li},
  \citenamefont {Moussa}, \citenamefont {Sounas},\ and\ \citenamefont
  {Al{\`u}}}]{li2020parity}%
  \BibitemOpen
  \bibfield  {author} {\bibinfo {author} {\bibfnamefont {H.}~\bibnamefont
  {Li}}, \bibinfo {author} {\bibfnamefont {H.}~\bibnamefont {Moussa}}, \bibinfo
  {author} {\bibfnamefont {D.}~\bibnamefont {Sounas}},\ and\ \bibinfo {author}
  {\bibfnamefont {A.}~\bibnamefont {Al{\`u}}},\ }\bibfield  {title} {\bibinfo
  {title} {Parity-time symmetry based on time modulation},\ }\href@noop {}
  {\bibfield  {journal} {\bibinfo  {journal} {Physical Review Applied}\
  }\textbf {\bibinfo {volume} {14}},\ \bibinfo {pages} {031002} (\bibinfo
  {year} {2020})}\BibitemShut {NoStop}%
\bibitem [{\citenamefont {Jiang}\ \emph {et~al.}(2021)\citenamefont {Jiang},
  \citenamefont {Su}, \citenamefont {Wu}, \citenamefont {Peng},\ and\
  \citenamefont {Budker}}]{jiang2021floquet}%
  \BibitemOpen
  \bibfield  {author} {\bibinfo {author} {\bibfnamefont {M.}~\bibnamefont
  {Jiang}}, \bibinfo {author} {\bibfnamefont {H.}~\bibnamefont {Su}}, \bibinfo
  {author} {\bibfnamefont {Z.}~\bibnamefont {Wu}}, \bibinfo {author}
  {\bibfnamefont {X.}~\bibnamefont {Peng}},\ and\ \bibinfo {author}
  {\bibfnamefont {D.}~\bibnamefont {Budker}},\ }\bibfield  {title} {\bibinfo
  {title} {Floquet maser},\ }\href@noop {} {\bibfield  {journal} {\bibinfo
  {journal} {Science Advances}\ }\textbf {\bibinfo {volume} {7}},\ \bibinfo
  {pages} {eabe0719} (\bibinfo {year} {2021})}\BibitemShut {NoStop}%
\bibitem [{\citenamefont {Jayathurathnage}\ \emph {et~al.}(2021)\citenamefont
  {Jayathurathnage}, \citenamefont {Liu}, \citenamefont {Mirmoosa},
  \citenamefont {Wang}, \citenamefont {Fleury},\ and\ \citenamefont
  {Tretyakov}}]{PhysRevApplied.16.014017}%
  \BibitemOpen
  \bibfield  {author} {\bibinfo {author} {\bibfnamefont {P.}~\bibnamefont
  {Jayathurathnage}}, \bibinfo {author} {\bibfnamefont {F.}~\bibnamefont
  {Liu}}, \bibinfo {author} {\bibfnamefont {M.~S.}\ \bibnamefont {Mirmoosa}},
  \bibinfo {author} {\bibfnamefont {X.}~\bibnamefont {Wang}}, \bibinfo {author}
  {\bibfnamefont {R.}~\bibnamefont {Fleury}},\ and\ \bibinfo {author}
  {\bibfnamefont {S.~A.}\ \bibnamefont {Tretyakov}},\ }\bibfield  {title}
  {\bibinfo {title} {Time-varying components for enhancing wireless transfer of
  power and information},\ }\href
  {https://doi.org/10.1103/PhysRevApplied.16.014017} {\bibfield  {journal}
  {\bibinfo  {journal} {Phys. Rev. Appl.}\ }\textbf {\bibinfo {volume} {16}},\
  \bibinfo {pages} {014017} (\bibinfo {year} {2021})}\BibitemShut {NoStop}%
\bibitem [{\citenamefont {Xu}\ \emph {et~al.}(2022{\natexlab{b}})\citenamefont
  {Xu}, \citenamefont {Wang}, \citenamefont {Dai}, \citenamefont {Mao},
  \citenamefont {Cai}, \citenamefont {Zhu},\ and\ \citenamefont
  {Wang}}]{PhysRevLett.129.273603}%
  \BibitemOpen
  \bibfield  {author} {\bibinfo {author} {\bibfnamefont {X.}~\bibnamefont
  {Xu}}, \bibinfo {author} {\bibfnamefont {J.}~\bibnamefont {Wang}}, \bibinfo
  {author} {\bibfnamefont {J.}~\bibnamefont {Dai}}, \bibinfo {author}
  {\bibfnamefont {R.}~\bibnamefont {Mao}}, \bibinfo {author} {\bibfnamefont
  {H.}~\bibnamefont {Cai}}, \bibinfo {author} {\bibfnamefont {S.-Y.}\
  \bibnamefont {Zhu}},\ and\ \bibinfo {author} {\bibfnamefont {D.-W.}\
  \bibnamefont {Wang}},\ }\bibfield  {title} {\bibinfo {title} {Floquet
  superradiance lattices in thermal atoms},\ }\href
  {https://doi.org/10.1103/PhysRevLett.129.273603} {\bibfield  {journal}
  {\bibinfo  {journal} {Phys. Rev. Lett.}\ }\textbf {\bibinfo {volume} {129}},\
  \bibinfo {pages} {273603} (\bibinfo {year} {2022}{\natexlab{b}})}\BibitemShut
  {NoStop}%
\bibitem [{\citenamefont {Block}\ \emph {et~al.}(2014)\citenamefont {Block},
  \citenamefont {Etrich}, \citenamefont {Limboeck}, \citenamefont {Bleckmann},
  \citenamefont {Soergel}, \citenamefont {Rockstuhl},\ and\ \citenamefont
  {Linden}}]{block2014bloch}%
  \BibitemOpen
  \bibfield  {author} {\bibinfo {author} {\bibfnamefont {A.}~\bibnamefont
  {Block}}, \bibinfo {author} {\bibfnamefont {C.}~\bibnamefont {Etrich}},
  \bibinfo {author} {\bibfnamefont {T.}~\bibnamefont {Limboeck}}, \bibinfo
  {author} {\bibfnamefont {F.}~\bibnamefont {Bleckmann}}, \bibinfo {author}
  {\bibfnamefont {E.}~\bibnamefont {Soergel}}, \bibinfo {author} {\bibfnamefont
  {C.}~\bibnamefont {Rockstuhl}},\ and\ \bibinfo {author} {\bibfnamefont
  {S.}~\bibnamefont {Linden}},\ }\bibfield  {title} {\bibinfo {title} {Bloch
  oscillations in plasmonic waveguide arrays},\ }\href@noop {} {\bibfield
  {journal} {\bibinfo  {journal} {Nature communications}\ }\textbf {\bibinfo
  {volume} {5}},\ \bibinfo {pages} {3843} (\bibinfo {year} {2014})}\BibitemShut
  {NoStop}%
\bibitem [{\citenamefont {Rechtsman}\ \emph {et~al.}(2013)\citenamefont
  {Rechtsman}, \citenamefont {Zeuner}, \citenamefont {Plotnik}, \citenamefont
  {Lumer}, \citenamefont {Podolsky}, \citenamefont {Dreisow}, \citenamefont
  {Nolte}, \citenamefont {Segev},\ and\ \citenamefont
  {Szameit}}]{rechtsman2013photonic}%
  \BibitemOpen
  \bibfield  {author} {\bibinfo {author} {\bibfnamefont {M.~C.}\ \bibnamefont
  {Rechtsman}}, \bibinfo {author} {\bibfnamefont {J.~M.}\ \bibnamefont
  {Zeuner}}, \bibinfo {author} {\bibfnamefont {Y.}~\bibnamefont {Plotnik}},
  \bibinfo {author} {\bibfnamefont {Y.}~\bibnamefont {Lumer}}, \bibinfo
  {author} {\bibfnamefont {D.}~\bibnamefont {Podolsky}}, \bibinfo {author}
  {\bibfnamefont {F.}~\bibnamefont {Dreisow}}, \bibinfo {author} {\bibfnamefont
  {S.}~\bibnamefont {Nolte}}, \bibinfo {author} {\bibfnamefont
  {M.}~\bibnamefont {Segev}},\ and\ \bibinfo {author} {\bibfnamefont
  {A.}~\bibnamefont {Szameit}},\ }\bibfield  {title} {\bibinfo {title}
  {Photonic floquet topological insulators},\ }\href@noop {} {\bibfield
  {journal} {\bibinfo  {journal} {Nature}\ }\textbf {\bibinfo {volume} {496}},\
  \bibinfo {pages} {196} (\bibinfo {year} {2013})}\BibitemShut {NoStop}%
\bibitem [{\citenamefont {Pan}\ and\ \citenamefont
  {Wang}(2020)}]{PhysRevResearch.2.043239}%
  \BibitemOpen
  \bibfield  {author} {\bibinfo {author} {\bibfnamefont {Y.}~\bibnamefont
  {Pan}}\ and\ \bibinfo {author} {\bibfnamefont {B.}~\bibnamefont {Wang}},\
  }\bibfield  {title} {\bibinfo {title} {Time-crystalline phases and
  period-doubling oscillations in one-dimensional floquet topological
  insulators},\ }\href {https://doi.org/10.1103/PhysRevResearch.2.043239}
  {\bibfield  {journal} {\bibinfo  {journal} {Phys. Rev. Res.}\ }\textbf
  {\bibinfo {volume} {2}},\ \bibinfo {pages} {043239} (\bibinfo {year}
  {2020})}\BibitemShut {NoStop}%
\bibitem [{\citenamefont {Lin}\ \emph {et~al.}(2023)\citenamefont {Lin},
  \citenamefont {Song}, \citenamefont {Sun}, \citenamefont {Li}, \citenamefont
  {Huang}, \citenamefont {Wu}, \citenamefont {Xin}, \citenamefont {Zhu},\ and\
  \citenamefont {Li}}]{PhysRevApplied.20.054005}%
  \BibitemOpen
  \bibfield  {author} {\bibinfo {author} {\bibfnamefont {Z.}~\bibnamefont
  {Lin}}, \bibinfo {author} {\bibfnamefont {W.}~\bibnamefont {Song}}, \bibinfo
  {author} {\bibfnamefont {J.}~\bibnamefont {Sun}}, \bibinfo {author}
  {\bibfnamefont {X.}~\bibnamefont {Li}}, \bibinfo {author} {\bibfnamefont
  {C.}~\bibnamefont {Huang}}, \bibinfo {author} {\bibfnamefont
  {S.}~\bibnamefont {Wu}}, \bibinfo {author} {\bibfnamefont {H.}~\bibnamefont
  {Xin}}, \bibinfo {author} {\bibfnamefont {S.}~\bibnamefont {Zhu}},\ and\
  \bibinfo {author} {\bibfnamefont {T.}~\bibnamefont {Li}},\ }\bibfield
  {title} {\bibinfo {title} {Ultrabroadband low-crosstalk dense lithium niobate
  waveguides by floquet engineering},\ }\href
  {https://doi.org/10.1103/PhysRevApplied.20.054005} {\bibfield  {journal}
  {\bibinfo  {journal} {Phys. Rev. Appl.}\ }\textbf {\bibinfo {volume} {20}},\
  \bibinfo {pages} {054005} (\bibinfo {year} {2023})}\BibitemShut {NoStop}%
\bibitem [{\citenamefont {Pyrialakos}\ \emph {et~al.}(2022)\citenamefont
  {Pyrialakos}, \citenamefont {Beck}, \citenamefont {Heinrich}, \citenamefont
  {Maczewsky}, \citenamefont {Kantartzis}, \citenamefont {Khajavikhan},
  \citenamefont {Szameit},\ and\ \citenamefont
  {Christodoulides}}]{pyrialakos2022bimorphic}%
  \BibitemOpen
  \bibfield  {author} {\bibinfo {author} {\bibfnamefont {G.~G.}\ \bibnamefont
  {Pyrialakos}}, \bibinfo {author} {\bibfnamefont {J.}~\bibnamefont {Beck}},
  \bibinfo {author} {\bibfnamefont {M.}~\bibnamefont {Heinrich}}, \bibinfo
  {author} {\bibfnamefont {L.~J.}\ \bibnamefont {Maczewsky}}, \bibinfo {author}
  {\bibfnamefont {N.~V.}\ \bibnamefont {Kantartzis}}, \bibinfo {author}
  {\bibfnamefont {M.}~\bibnamefont {Khajavikhan}}, \bibinfo {author}
  {\bibfnamefont {A.}~\bibnamefont {Szameit}},\ and\ \bibinfo {author}
  {\bibfnamefont {D.~N.}\ \bibnamefont {Christodoulides}},\ }\bibfield  {title}
  {\bibinfo {title} {Bimorphic floquet topological insulators},\ }\href@noop {}
  {\bibfield  {journal} {\bibinfo  {journal} {Nature Materials}\ }\textbf
  {\bibinfo {volume} {21}},\ \bibinfo {pages} {634} (\bibinfo {year}
  {2022})}\BibitemShut {NoStop}%
\bibitem [{\citenamefont {Lin}\ \emph {et~al.}(2024)\citenamefont {Lin},
  \citenamefont {Song}, \citenamefont {Wang}, \citenamefont {Xin},
  \citenamefont {Sun}, \citenamefont {Wu}, \citenamefont {Huang}, \citenamefont
  {Zhu}, \citenamefont {Jiang},\ and\ \citenamefont
  {Li}}]{PhysRevLett.133.073803}%
  \BibitemOpen
  \bibfield  {author} {\bibinfo {author} {\bibfnamefont {Z.}~\bibnamefont
  {Lin}}, \bibinfo {author} {\bibfnamefont {W.}~\bibnamefont {Song}}, \bibinfo
  {author} {\bibfnamefont {L.-W.}\ \bibnamefont {Wang}}, \bibinfo {author}
  {\bibfnamefont {H.}~\bibnamefont {Xin}}, \bibinfo {author} {\bibfnamefont
  {J.}~\bibnamefont {Sun}}, \bibinfo {author} {\bibfnamefont {S.}~\bibnamefont
  {Wu}}, \bibinfo {author} {\bibfnamefont {C.}~\bibnamefont {Huang}}, \bibinfo
  {author} {\bibfnamefont {S.}~\bibnamefont {Zhu}}, \bibinfo {author}
  {\bibfnamefont {J.-H.}\ \bibnamefont {Jiang}},\ and\ \bibinfo {author}
  {\bibfnamefont {T.}~\bibnamefont {Li}},\ }\bibfield  {title} {\bibinfo
  {title} {Observation of topological transition in floquet non-hermitian skin
  effects in silicon photonics},\ }\href
  {https://doi.org/10.1103/PhysRevLett.133.073803} {\bibfield  {journal}
  {\bibinfo  {journal} {Phys. Rev. Lett.}\ }\textbf {\bibinfo {volume} {133}},\
  \bibinfo {pages} {073803} (\bibinfo {year} {2024})}\BibitemShut {NoStop}%
\bibitem [{\citenamefont {Fedorova}\ \emph {et~al.}(2019)\citenamefont
  {Fedorova}, \citenamefont {J{\"o}rg}, \citenamefont {Dauer}, \citenamefont
  {Letscher}, \citenamefont {Fleischhauer}, \citenamefont {Eggert},
  \citenamefont {Linden},\ and\ \citenamefont {von
  Freymann}}]{fedorova2019limits}%
  \BibitemOpen
  \bibfield  {author} {\bibinfo {author} {\bibfnamefont {Z.}~\bibnamefont
  {Fedorova}}, \bibinfo {author} {\bibfnamefont {C.}~\bibnamefont {J{\"o}rg}},
  \bibinfo {author} {\bibfnamefont {C.}~\bibnamefont {Dauer}}, \bibinfo
  {author} {\bibfnamefont {F.}~\bibnamefont {Letscher}}, \bibinfo {author}
  {\bibfnamefont {M.}~\bibnamefont {Fleischhauer}}, \bibinfo {author}
  {\bibfnamefont {S.}~\bibnamefont {Eggert}}, \bibinfo {author} {\bibfnamefont
  {S.}~\bibnamefont {Linden}},\ and\ \bibinfo {author} {\bibfnamefont
  {G.}~\bibnamefont {von Freymann}},\ }\bibfield  {title} {\bibinfo {title}
  {Limits of topological protection under local periodic driving},\ }\href@noop
  {} {\bibfield  {journal} {\bibinfo  {journal} {Light: Science \&
  Applications}\ }\textbf {\bibinfo {volume} {8}},\ \bibinfo {pages} {63}
  (\bibinfo {year} {2019})}\BibitemShut {NoStop}%
\bibitem [{\citenamefont {Ma}\ \emph {et~al.}(2018)\citenamefont {Ma},
  \citenamefont {Wang},\ and\ \citenamefont {An}}]{PhysRevA.97.023808}%
  \BibitemOpen
  \bibfield  {author} {\bibinfo {author} {\bibfnamefont {C.}~\bibnamefont
  {Ma}}, \bibinfo {author} {\bibfnamefont {Y.-S.}\ \bibnamefont {Wang}},\ and\
  \bibinfo {author} {\bibfnamefont {J.-H.}\ \bibnamefont {An}},\ }\bibfield
  {title} {\bibinfo {title} {Floquet engineering of localized propagation of
  light in a waveguide array},\ }\href
  {https://doi.org/10.1103/PhysRevA.97.023808} {\bibfield  {journal} {\bibinfo
  {journal} {Phys. Rev. A}\ }\textbf {\bibinfo {volume} {97}},\ \bibinfo
  {pages} {023808} (\bibinfo {year} {2018})}\BibitemShut {NoStop}%
\bibitem [{\citenamefont {Lumer}\ \emph {et~al.}(2019)\citenamefont {Lumer},
  \citenamefont {Bandres}, \citenamefont {Heinrich}, \citenamefont {Maczewsky},
  \citenamefont {Herzig-Sheinfux}, \citenamefont {Szameit},\ and\ \citenamefont
  {Segev}}]{lumer2019light}%
  \BibitemOpen
  \bibfield  {author} {\bibinfo {author} {\bibfnamefont {Y.}~\bibnamefont
  {Lumer}}, \bibinfo {author} {\bibfnamefont {M.~A.}\ \bibnamefont {Bandres}},
  \bibinfo {author} {\bibfnamefont {M.}~\bibnamefont {Heinrich}}, \bibinfo
  {author} {\bibfnamefont {L.~J.}\ \bibnamefont {Maczewsky}}, \bibinfo {author}
  {\bibfnamefont {H.}~\bibnamefont {Herzig-Sheinfux}}, \bibinfo {author}
  {\bibfnamefont {A.}~\bibnamefont {Szameit}},\ and\ \bibinfo {author}
  {\bibfnamefont {M.}~\bibnamefont {Segev}},\ }\bibfield  {title} {\bibinfo
  {title} {Light guiding by artificial gauge fields},\ }\href@noop {}
  {\bibfield  {journal} {\bibinfo  {journal} {Nature Photonics}\ }\textbf
  {\bibinfo {volume} {13}},\ \bibinfo {pages} {339} (\bibinfo {year}
  {2019})}\BibitemShut {NoStop}%
\bibitem [{\citenamefont {Ivanov}\ \emph {et~al.}(2021)\citenamefont {Ivanov},
  \citenamefont {Kartashov},\ and\ \citenamefont
  {Konotop}}]{ivanov2021floquet}%
  \BibitemOpen
  \bibfield  {author} {\bibinfo {author} {\bibfnamefont {S.~K.}\ \bibnamefont
  {Ivanov}}, \bibinfo {author} {\bibfnamefont {Y.~V.}\ \bibnamefont
  {Kartashov}},\ and\ \bibinfo {author} {\bibfnamefont {V.~V.}\ \bibnamefont
  {Konotop}},\ }\bibfield  {title} {\bibinfo {title} {Floquet defect
  solitons},\ }\href@noop {} {\bibfield  {journal} {\bibinfo  {journal} {Optics
  Letters}\ }\textbf {\bibinfo {volume} {46}},\ \bibinfo {pages} {5364}
  (\bibinfo {year} {2021})}\BibitemShut {NoStop}%
\bibitem [{\citenamefont {He}\ \emph {et~al.}(2019)\citenamefont {He},
  \citenamefont {Addison}, \citenamefont {Jin}, \citenamefont {Mele},
  \citenamefont {Johnson},\ and\ \citenamefont {Zhen}}]{he2019floquet}%
  \BibitemOpen
  \bibfield  {author} {\bibinfo {author} {\bibfnamefont {L.}~\bibnamefont
  {He}}, \bibinfo {author} {\bibfnamefont {Z.}~\bibnamefont {Addison}},
  \bibinfo {author} {\bibfnamefont {J.}~\bibnamefont {Jin}}, \bibinfo {author}
  {\bibfnamefont {E.~J.}\ \bibnamefont {Mele}}, \bibinfo {author}
  {\bibfnamefont {S.~G.}\ \bibnamefont {Johnson}},\ and\ \bibinfo {author}
  {\bibfnamefont {B.}~\bibnamefont {Zhen}},\ }\bibfield  {title} {\bibinfo
  {title} {Floquet chern insulators of light},\ }\href@noop {} {\bibfield
  {journal} {\bibinfo  {journal} {Nature communications}\ }\textbf {\bibinfo
  {volume} {10}},\ \bibinfo {pages} {4194} (\bibinfo {year}
  {2019})}\BibitemShut {NoStop}%
\bibitem [{\citenamefont {Cheng}\ \emph {et~al.}(2019)\citenamefont {Cheng},
  \citenamefont {Pan}, \citenamefont {Wang}, \citenamefont {Zhang},
  \citenamefont {Yu}, \citenamefont {Gover}, \citenamefont {Zhang},
  \citenamefont {Li}, \citenamefont {Zhou},\ and\ \citenamefont
  {Zhu}}]{PhysRevLett.122.173901}%
  \BibitemOpen
  \bibfield  {author} {\bibinfo {author} {\bibfnamefont {Q.}~\bibnamefont
  {Cheng}}, \bibinfo {author} {\bibfnamefont {Y.}~\bibnamefont {Pan}}, \bibinfo
  {author} {\bibfnamefont {H.}~\bibnamefont {Wang}}, \bibinfo {author}
  {\bibfnamefont {C.}~\bibnamefont {Zhang}}, \bibinfo {author} {\bibfnamefont
  {D.}~\bibnamefont {Yu}}, \bibinfo {author} {\bibfnamefont {A.}~\bibnamefont
  {Gover}}, \bibinfo {author} {\bibfnamefont {H.}~\bibnamefont {Zhang}},
  \bibinfo {author} {\bibfnamefont {T.}~\bibnamefont {Li}}, \bibinfo {author}
  {\bibfnamefont {L.}~\bibnamefont {Zhou}},\ and\ \bibinfo {author}
  {\bibfnamefont {S.}~\bibnamefont {Zhu}},\ }\bibfield  {title} {\bibinfo
  {title} {Observation of anomalous $\ensuremath{\pi}$ modes in photonic
  floquet engineering},\ }\href
  {https://doi.org/10.1103/PhysRevLett.122.173901} {\bibfield  {journal}
  {\bibinfo  {journal} {Phys. Rev. Lett.}\ }\textbf {\bibinfo {volume} {122}},\
  \bibinfo {pages} {173901} (\bibinfo {year} {2019})}\BibitemShut {NoStop}%
\bibitem [{\citenamefont {Weidemann}\ \emph {et~al.}(2022)\citenamefont
  {Weidemann}, \citenamefont {Kremer}, \citenamefont {Longhi},\ and\
  \citenamefont {Szameit}}]{weidemann2022topological}%
  \BibitemOpen
  \bibfield  {author} {\bibinfo {author} {\bibfnamefont {S.}~\bibnamefont
  {Weidemann}}, \bibinfo {author} {\bibfnamefont {M.}~\bibnamefont {Kremer}},
  \bibinfo {author} {\bibfnamefont {S.}~\bibnamefont {Longhi}},\ and\ \bibinfo
  {author} {\bibfnamefont {A.}~\bibnamefont {Szameit}},\ }\bibfield  {title}
  {\bibinfo {title} {Topological triple phase transition in non-hermitian
  floquet quasicrystals},\ }\href@noop {} {\bibfield  {journal} {\bibinfo
  {journal} {Nature}\ }\textbf {\bibinfo {volume} {601}},\ \bibinfo {pages}
  {354} (\bibinfo {year} {2022})}\BibitemShut {NoStop}%
\bibitem [{\citenamefont {Park}\ \emph {et~al.}(2022)\citenamefont {Park},
  \citenamefont {Cho}, \citenamefont {Lee}, \citenamefont {Lee}, \citenamefont
  {Lee}, \citenamefont {Park}, \citenamefont {Ryu}, \citenamefont {Park},
  \citenamefont {Jeon},\ and\ \citenamefont {Min}}]{park2022revealing}%
  \BibitemOpen
  \bibfield  {author} {\bibinfo {author} {\bibfnamefont {J.}~\bibnamefont
  {Park}}, \bibinfo {author} {\bibfnamefont {H.}~\bibnamefont {Cho}}, \bibinfo
  {author} {\bibfnamefont {S.}~\bibnamefont {Lee}}, \bibinfo {author}
  {\bibfnamefont {K.}~\bibnamefont {Lee}}, \bibinfo {author} {\bibfnamefont
  {K.}~\bibnamefont {Lee}}, \bibinfo {author} {\bibfnamefont {H.~C.}\
  \bibnamefont {Park}}, \bibinfo {author} {\bibfnamefont {J.-W.}\ \bibnamefont
  {Ryu}}, \bibinfo {author} {\bibfnamefont {N.}~\bibnamefont {Park}}, \bibinfo
  {author} {\bibfnamefont {S.}~\bibnamefont {Jeon}},\ and\ \bibinfo {author}
  {\bibfnamefont {B.}~\bibnamefont {Min}},\ }\bibfield  {title} {\bibinfo
  {title} {Revealing non-hermitian band structure of photonic floquet media},\
  }\href@noop {} {\bibfield  {journal} {\bibinfo  {journal} {Science advances}\
  }\textbf {\bibinfo {volume} {8}},\ \bibinfo {pages} {eabo6220} (\bibinfo
  {year} {2022})}\BibitemShut {NoStop}%
\bibitem [{\citenamefont {Gandhi}\ and\ \citenamefont
  {Bandyopadhyay}(2023)}]{PhysRevB.108.014204}%
  \BibitemOpen
  \bibfield  {author} {\bibinfo {author} {\bibfnamefont {S.}~\bibnamefont
  {Gandhi}}\ and\ \bibinfo {author} {\bibfnamefont {J.~N.}\ \bibnamefont
  {Bandyopadhyay}},\ }\bibfield  {title} {\bibinfo {title} {Topological triple
  phase transition in non-hermitian quasicrystals with complex asymmetric
  hopping},\ }\href {https://doi.org/10.1103/PhysRevB.108.014204} {\bibfield
  {journal} {\bibinfo  {journal} {Phys. Rev. B}\ }\textbf {\bibinfo {volume}
  {108}},\ \bibinfo {pages} {014204} (\bibinfo {year} {2023})}\BibitemShut
  {NoStop}%
\bibitem [{\citenamefont {Zhou}\ \emph {et~al.}(2010)\citenamefont {Zhou},
  \citenamefont {Guo}, \citenamefont {Wang},\ and\ \citenamefont
  {Liu}}]{Zhou:10}%
  \BibitemOpen
  \bibfield  {author} {\bibinfo {author} {\bibfnamefont {K.}~\bibnamefont
  {Zhou}}, \bibinfo {author} {\bibfnamefont {Z.}~\bibnamefont {Guo}}, \bibinfo
  {author} {\bibfnamefont {J.}~\bibnamefont {Wang}},\ and\ \bibinfo {author}
  {\bibfnamefont {S.}~\bibnamefont {Liu}},\ }\bibfield  {title} {\bibinfo
  {title} {Defect modes in defective parity-time symmetric periodic complex
  potentials},\ }\href {https://doi.org/10.1364/OL.35.002928} {\bibfield
  {journal} {\bibinfo  {journal} {Opt. Lett.}\ }\textbf {\bibinfo {volume}
  {35}},\ \bibinfo {pages} {2928} (\bibinfo {year} {2010})}\BibitemShut
  {NoStop}%
\bibitem [{\citenamefont {Wang}\ and\ \citenamefont {Wang}(2011)}]{Wang:11}%
  \BibitemOpen
  \bibfield  {author} {\bibinfo {author} {\bibfnamefont {H.}~\bibnamefont
  {Wang}}\ and\ \bibinfo {author} {\bibfnamefont {J.}~\bibnamefont {Wang}},\
  }\bibfield  {title} {\bibinfo {title} {Defect solitons in parity-time
  periodic potentials},\ }\href {https://doi.org/10.1364/OE.19.004030}
  {\bibfield  {journal} {\bibinfo  {journal} {Opt. Express}\ }\textbf {\bibinfo
  {volume} {19}},\ \bibinfo {pages} {4030} (\bibinfo {year}
  {2011})}\BibitemShut {NoStop}%
\bibitem [{\citenamefont {Makris}\ \emph {et~al.}(2008)\citenamefont {Makris},
  \citenamefont {El-Ganainy}, \citenamefont {Christodoulides},\ and\
  \citenamefont {Musslimani}}]{PhysRevLett.100.103904}%
  \BibitemOpen
  \bibfield  {author} {\bibinfo {author} {\bibfnamefont {K.~G.}\ \bibnamefont
  {Makris}}, \bibinfo {author} {\bibfnamefont {R.}~\bibnamefont {El-Ganainy}},
  \bibinfo {author} {\bibfnamefont {D.~N.}\ \bibnamefont {Christodoulides}},\
  and\ \bibinfo {author} {\bibfnamefont {Z.~H.}\ \bibnamefont {Musslimani}},\
  }\bibfield  {title} {\bibinfo {title} {Beam dynamics in
  $\mathcal{P}\mathcal{T}$ symmetric optical lattices},\ }\href
  {https://doi.org/10.1103/PhysRevLett.100.103904} {\bibfield  {journal}
  {\bibinfo  {journal} {Phys. Rev. Lett.}\ }\textbf {\bibinfo {volume} {100}},\
  \bibinfo {pages} {103904} (\bibinfo {year} {2008})}\BibitemShut {NoStop}%
\bibitem [{\citenamefont {Dehghani}\ \emph {et~al.}(2014)\citenamefont
  {Dehghani}, \citenamefont {Oka},\ and\ \citenamefont
  {Mitra}}]{PhysRevB.90.195429}%
  \BibitemOpen
  \bibfield  {author} {\bibinfo {author} {\bibfnamefont {H.}~\bibnamefont
  {Dehghani}}, \bibinfo {author} {\bibfnamefont {T.}~\bibnamefont {Oka}},\ and\
  \bibinfo {author} {\bibfnamefont {A.}~\bibnamefont {Mitra}},\ }\bibfield
  {title} {\bibinfo {title} {Dissipative floquet topological systems},\ }\href
  {https://doi.org/10.1103/PhysRevB.90.195429} {\bibfield  {journal} {\bibinfo
  {journal} {Phys. Rev. B}\ }\textbf {\bibinfo {volume} {90}},\ \bibinfo
  {pages} {195429} (\bibinfo {year} {2014})}\BibitemShut {NoStop}%
\bibitem [{\citenamefont {Wu}\ \emph {et~al.}(2021)\citenamefont {Wu},
  \citenamefont {Song}, \citenamefont {Gao}, \citenamefont {Chen},
  \citenamefont {Zhu},\ and\ \citenamefont {Li}}]{PhysRevResearch.3.023211}%
  \BibitemOpen
  \bibfield  {author} {\bibinfo {author} {\bibfnamefont {S.}~\bibnamefont
  {Wu}}, \bibinfo {author} {\bibfnamefont {W.}~\bibnamefont {Song}}, \bibinfo
  {author} {\bibfnamefont {S.}~\bibnamefont {Gao}}, \bibinfo {author}
  {\bibfnamefont {Y.}~\bibnamefont {Chen}}, \bibinfo {author} {\bibfnamefont
  {S.}~\bibnamefont {Zhu}},\ and\ \bibinfo {author} {\bibfnamefont
  {T.}~\bibnamefont {Li}},\ }\bibfield  {title} {\bibinfo {title} {Floquet
  $\ensuremath{\pi}$ mode engineering in non-hermitian waveguide lattices},\
  }\href {https://doi.org/10.1103/PhysRevResearch.3.023211} {\bibfield
  {journal} {\bibinfo  {journal} {Phys. Rev. Res.}\ }\textbf {\bibinfo {volume}
  {3}},\ \bibinfo {pages} {023211} (\bibinfo {year} {2021})}\BibitemShut
  {NoStop}%
\bibitem [{\citenamefont {Zhen}\ \emph {et~al.}(2015)\citenamefont {Zhen},
  \citenamefont {Hsu}, \citenamefont {Igarashi}, \citenamefont {Lu},
  \citenamefont {Kaminer}, \citenamefont {Pick}, \citenamefont {Chua},
  \citenamefont {Joannopoulos},\ and\ \citenamefont
  {Solja{\v{c}}i{\'c}}}]{zhen2015spawning}%
  \BibitemOpen
  \bibfield  {author} {\bibinfo {author} {\bibfnamefont {B.}~\bibnamefont
  {Zhen}}, \bibinfo {author} {\bibfnamefont {C.~W.}\ \bibnamefont {Hsu}},
  \bibinfo {author} {\bibfnamefont {Y.}~\bibnamefont {Igarashi}}, \bibinfo
  {author} {\bibfnamefont {L.}~\bibnamefont {Lu}}, \bibinfo {author}
  {\bibfnamefont {I.}~\bibnamefont {Kaminer}}, \bibinfo {author} {\bibfnamefont
  {A.}~\bibnamefont {Pick}}, \bibinfo {author} {\bibfnamefont {S.-L.}\
  \bibnamefont {Chua}}, \bibinfo {author} {\bibfnamefont {J.~D.}\ \bibnamefont
  {Joannopoulos}},\ and\ \bibinfo {author} {\bibfnamefont {M.}~\bibnamefont
  {Solja{\v{c}}i{\'c}}},\ }\bibfield  {title} {\bibinfo {title} {Spawning rings
  of exceptional points out of dirac cones},\ }\href@noop {} {\bibfield
  {journal} {\bibinfo  {journal} {Nature}\ }\textbf {\bibinfo {volume} {525}},\
  \bibinfo {pages} {354} (\bibinfo {year} {2015})}\BibitemShut {NoStop}%
\bibitem [{\citenamefont {Feng}\ \emph {et~al.}(2023)\citenamefont {Feng},
  \citenamefont {Liu}, \citenamefont {Liu}, \citenamefont {Yu}, \citenamefont
  {Liang}, \citenamefont {Li}, \citenamefont {Zhang}, \citenamefont {Xiao},\
  and\ \citenamefont {Zhang}}]{PhysRevLett.131.013802}%
  \BibitemOpen
  \bibfield  {author} {\bibinfo {author} {\bibfnamefont {Y.}~\bibnamefont
  {Feng}}, \bibinfo {author} {\bibfnamefont {Z.}~\bibnamefont {Liu}}, \bibinfo
  {author} {\bibfnamefont {F.}~\bibnamefont {Liu}}, \bibinfo {author}
  {\bibfnamefont {J.}~\bibnamefont {Yu}}, \bibinfo {author} {\bibfnamefont
  {S.}~\bibnamefont {Liang}}, \bibinfo {author} {\bibfnamefont
  {F.}~\bibnamefont {Li}}, \bibinfo {author} {\bibfnamefont {Y.}~\bibnamefont
  {Zhang}}, \bibinfo {author} {\bibfnamefont {M.}~\bibnamefont {Xiao}},\ and\
  \bibinfo {author} {\bibfnamefont {Z.}~\bibnamefont {Zhang}},\ }\bibfield
  {title} {\bibinfo {title} {Loss difference induced localization in a
  non-hermitian honeycomb photonic lattice},\ }\href
  {https://doi.org/10.1103/PhysRevLett.131.013802} {\bibfield  {journal}
  {\bibinfo  {journal} {Phys. Rev. Lett.}\ }\textbf {\bibinfo {volume} {131}},\
  \bibinfo {pages} {013802} (\bibinfo {year} {2023})}\BibitemShut {NoStop}%
\bibitem [{\citenamefont {Qi}\ \emph {et~al.}(2018)\citenamefont {Qi},
  \citenamefont {Zhang},\ and\ \citenamefont {Ge}}]{PhysRevLett.120.093901}%
  \BibitemOpen
  \bibfield  {author} {\bibinfo {author} {\bibfnamefont {B.}~\bibnamefont
  {Qi}}, \bibinfo {author} {\bibfnamefont {L.}~\bibnamefont {Zhang}},\ and\
  \bibinfo {author} {\bibfnamefont {L.}~\bibnamefont {Ge}},\ }\bibfield
  {title} {\bibinfo {title} {Defect states emerging from a non-hermitian
  flatband of photonic zero modes},\ }\href
  {https://doi.org/10.1103/PhysRevLett.120.093901} {\bibfield  {journal}
  {\bibinfo  {journal} {Phys. Rev. Lett.}\ }\textbf {\bibinfo {volume} {120}},\
  \bibinfo {pages} {093901} (\bibinfo {year} {2018})}\BibitemShut {NoStop}%
\bibitem [{\citenamefont {Ge}(2018)}]{ge2018non}%
  \BibitemOpen
  \bibfield  {author} {\bibinfo {author} {\bibfnamefont {L.}~\bibnamefont
  {Ge}},\ }\bibfield  {title} {\bibinfo {title} {Non-hermitian lattices with a
  flat band and polynomial power increase},\ }\href@noop {} {\bibfield
  {journal} {\bibinfo  {journal} {Photonics Research}\ }\textbf {\bibinfo
  {volume} {6}},\ \bibinfo {pages} {A10} (\bibinfo {year} {2018})}\BibitemShut
  {NoStop}%
\bibitem [{\citenamefont {Christodoulides}\ \emph {et~al.}(2003)\citenamefont
  {Christodoulides}, \citenamefont {Lederer},\ and\ \citenamefont
  {Silberberg}}]{christodoulides2003discretizing}%
  \BibitemOpen
  \bibfield  {author} {\bibinfo {author} {\bibfnamefont {D.~N.}\ \bibnamefont
  {Christodoulides}}, \bibinfo {author} {\bibfnamefont {F.}~\bibnamefont
  {Lederer}},\ and\ \bibinfo {author} {\bibfnamefont {Y.}~\bibnamefont
  {Silberberg}},\ }\bibfield  {title} {\bibinfo {title} {Discretizing light
  behaviour in linear and nonlinear waveguide lattices},\ }\href@noop {}
  {\bibfield  {journal} {\bibinfo  {journal} {Nature}\ }\textbf {\bibinfo
  {volume} {424}},\ \bibinfo {pages} {817} (\bibinfo {year}
  {2003})}\BibitemShut {NoStop}%
\end{thebibliography}%

\end{document}